\documentclass[preprint]{aastex}
\usepackage{color}

\newcommand{\Spitzer}{{\sl Spitzer}}

\newcommand{\HST}{{\sl HST}}

\newcommand{\Msun}{\mbox{$M_{\sun}$}}

\newcommand{\Lsun}{\mbox{$L_{\sun}$}}
\newcommand{\Rsun}{\mbox{$R_{\sun}$}}
\newcommand{\Mjup}{\mbox{$M_{\rm Jup}$}}
\newcommand{\Rjup}{\mbox{$R_{\rm Jup}$}}

\newcommand{\etal}{et al.}
\newcommand{\eg}{e.g.}
\newcommand{\ie}{i.e.}
\newcommand{\cf}{cf.}

\newcommand{\kms}{\hbox{km~s$^{-1}$}}
\newcommand{\rchisq}{\mbox{$\tilde\chi^2$}}

\newcommand{\htwoo}{{\hbox{H$_2$O}}}   
\newcommand{\htwo}{{\hbox{H$_2$}}}     
\newcommand{\meth}{{\hbox{CH$_4$}}}   
\newcommand{\Kp}{\mbox{$K^{\prime}$}}
\newcommand{\Ks}{\mbox{$K_S$}}
\newcommand{\degs}{\mbox{$^{\circ}$}}
\newcommand{\Lbol}{\mbox{$L_{\rm bol}$}}

\newcommand{\Teff}{\mbox{$T_{\rm eff}$}}
\newcommand{\logg}{\mbox{$\log(g)$}}

\newcommand{\Lp}{\mbox{${L^\prime}$}}

\newcommand{\eInd}{\hbox{$\epsilon$~Ind}}

\newcommand{\twomassbin}{\hbox{2MASS~J1534$-$2952AB}}

\slugcomment{ApJ, in press, 06/30/08}

\shorttitle{Dynamical Mass of a Binary T Dwarf}
\shortauthors{Liu \etal}

\begin{document}

\title{Keck Laser Guide Star Adaptive Optics Monitoring of
  2MASS~J15344984$-$2952274AB: First Dynamical Mass Determination of a
  Binary T~Dwarf \altaffilmark{1,2}}


\author{Michael C. Liu,\altaffilmark{3,4,5}
        Trent J. Dupuy,\altaffilmark{3}
        Michael J. Ireland\altaffilmark{6,7}}

\altaffiltext{1}{Most of the data presented herein were obtained at the
  W.M. Keck Observatory, which is operated as a scientific partnership
  among the California Institute of Technology, the University of
  California, and the National Aeronautics and Space Administration. The
  Observatory was made possible by the generous financial support of the
  W.M. Keck Foundation.}
\altaffiltext{2}{Based on observations made with the NASA/ESA
  Hubble Space Telescope, obtained from the Data Archive at the
  Space Telescope Science Institute, which is operated by the
  Association of Universities for Research in Astronomy, Inc., under
  NASA contract NAS 5-26555.}
\altaffiltext{3}{Institute for Astronomy, University of Hawai`i, 2680
  Woodlawn Drive, Honolulu, HI 96822; mliu@ifa.hawaii.edu}
\altaffiltext{4}{Alfred P. Sloan Research Fellow}
\altaffiltext{5}{Visiting Astronomer at the Infrared Telescope
  Facility, which is operated by the University of Hawaii under
  Cooperative Agreement no. NNX08AE38A with the National Aeronautics
  and Space Administration, Science Mission Directorate, Planetary
  Astronomy Program.}
\altaffiltext{6}{Division of Geological and Planetary Sciences, 
  California Institute of Technology, Pasadena, CA}
\altaffiltext{7}{School of Physics, University of Sydney, NSW 2006,
  Australia}

\begin{abstract} We present multi-epoch, near-infrared imaging of the
  binary T5.0+T5.5~dwarf 2MASS~J15344984$-$2952274AB obtained with the
  Keck laser guide star adaptive optics system.  Our Keck data achieve
  sub-milliarcsecond relative astrometry and combined with an
  extensive (re-)analysis of archival \HST\ imaging, the total dataset
  spans $\sim$50\% the orbital period.  We use a Markov Chain Monte
  Carlo analysis to determine an orbital period of
  15.1$_{-1.6}^{+3.1}$~yr and a semi-major axis of
  2.3$_{-0.2}^{+0.3}$~AU.  We measure a total mass of $0.056 \pm
  0.003$~\Msun\ ($59\pm3$~\Mjup), where the largest uncertainty arises
  from the parallax.  This is the first field binary for which both
  components are confirmed to be substellar.  This is also the coolest
  and lowest mass binary with a dynamical mass determination to date.
  Using evolutionary models and accounting for the measurement
  covariances, we derive an age of \hbox{0.78$\pm$0.09~Gyr} and a mass
  ratio of $0.936_{-0.008}^{+0.012}$.  The relatively youthful age is
  consistent with the low tangential velocity of this system.  For the
  individual components, we find \hbox{$\Teff = 1028\pm17$~K} and
  $978\pm17$~K, $\logg = 4.91\pm0.04$ and $4.87\pm0.04$ (cgs), and
  masses of $0.0287\pm0.0016~\Msun$ ($30.1\pm1.7$~\Mjup) and
  $0.0269\pm0.0016~\Msun$ ($28.2\pm1.7$~\Mjup).  These precise values
  generally agree with previous studies of T~dwarfs and affirm current
  theoretical models.  However, (1) the temperatures are about 100~K
  cooler than derived for similar field objects and suggest that the
  representative ages of field brown dwarfs may be overestimated.
  Similarly, (2) the H-R diagram positions are discrepant with current
  model predictions and taken at face value would overestimate the
  masses.  While this may arise from large errors in the luminosities
  and/or radii predicted by evolutionary models, the likely cause is a
  modest ($\approx$100~K) overestimate in temperature of T~dwarfs
  determined from model atmospheres.  We elucidate future tests of
  theory as the sample of substellar dynamical masses increases.  In
  particular, we suggest that field brown dwarf binaries with
  dynamical masses (``mass benchmarks'') can serve as reference points
  for \Teff\ and \logg\ and thereby constrain ultracool atmosphere
  models, as good as or even better than single brown dwarfs with age
  estimates (``age benchmarks'').
\end{abstract}



\section{Introduction}

Over about the past decade, the parameter space of traditional stellar
astrophysics has been greatly expanded with the discovery and
characterization of brown dwarfs, objects that for most of their
lifetimes are colder and less luminous than main-sequence stars.
Despite ample progress in finding and characterizing brown dwarfs,
very few direct measurements of their physical properties have been
made so far.  In particular, dynamical masses for brown dwarfs are
sorely needed to test the theoretical models over a wide range of
parameter space.  In comparison to the $>$100 binary stars with direct
mass determinations, dynamical masses have been measured for only a
handful of objects clearly below the stellar/substellar boundary:
\begin{enumerate}
\item[1.] the secondary component of the M8.5+M9 visual binary
  Gl~569Bab, which itself is a companion to a young
  ($\approx$100--300~Myr) field M2.5~dwarf \citep{2000ApJ...529L..37M,
    2001ApJ...560..390L, 2001ApJ...554L..67K, 2004astro.ph..7334O};
\item[2.+3.] the very young ($\sim$Myr) eclipsing M6.5+M6.5 binary
  brown dwarf 2MASS~J05352184$-$0546085 \citep{2006Natur.440..311S};
\item[4.] the secondary component of the GJ~802AB system, where the
  secondary has an estimated spectral type of L5 and the primary is a
  M5.5+M5.5 spectroscopic binary \citep{2006ApJ...649..389P,
    2006ApJ...650L.131L, gl802b-ireland}.
\end{enumerate}
In addition, GJ~569Ba itself may be a unresolved binary brown dwarf
\citep{2006ApJ...644.1183S}, and the secondary component of the
L0+L1.5 binary 2MASSW~J0746425+2000321AB, which appears to be an old
($\gtrsim$1~Gyr) field system, has a mass near the stellar/substellar
boundary \citep{2001AJ....121..489R, 2004A&A...423..341B,
  2006AJ....131..638G}.

About 100~ultracool visual binaries are known,\footnote{We follow the
  popular convention that ``ultracool'' refers to objects of
  (integrated-light) spectral type M6 or later.} found with high
angular resolution imaging surveys conducted by {\sl Hubble Space
  Telescope} (\HST; \eg, \citealp{2001AJ....121..489R,
  2003AJ....126.1526B, 2003AJ....125.3302G, 2003ApJ...586..512B,
  2006ApJS..166..585B}) and ground-based adaptive optics (AO) imaging
(\eg, \citealp{2003ApJ...587..407C, 2005AJ....129.2849B,
  2006astro.ph..5037L, 2008AJ....135..580R}; Liu \etal, in prep).
Only $\approx$10\% of these are binaries composed of the coldest class
of brown dwarf, the T~dwarfs.  T~dwarfs are distinguished by their
very red optical colors arising from pressure-broadened alkali
resonance lines and very blue near-infrared (IR) colors from strong
\meth, \htwoo, and collision-induced \htwo\ absorption
\citep[e.g.][]{1995Sci...270.1478O, geb01, kirk05}.
These are the lowest luminosity and coolest objects directly detected
outside of our solar system, with bolometric luminosities (\Lbol) of
$\lesssim10^{-4.5}\Lsun$ and effective temperatures (\Teff) of
$\approx700-1300$~K \citep[e.g.][]{2004AJ....127.2948V, gol04,
  2006ApJ...639.1095B, 2007arXiv0705.2602L, 2007MNRAS.381.1400W,
  2008A&A...482..961D}.  As such, analyzing their physical properties
is an important pathway to understanding the properties of gas-giant
extrasolar planets.

The subject of this paper is the T~dwarf 2MASS~J15344984$-$2952274AB,
hereinafter \twomassbin, which has an integrated-light infrared
spectral type of T5 \citep{burg01,
  2005astro.ph.10090B}.\footnote{\citet{2005astro.ph.10090B} report
  integrated-light spectral types of both T5 and T5.5 for \twomassbin\
  based on a spectrum obtained with the CTIO/OSIRIS instrument.
  Examination of their spectrum by us finds that T5 is the correct
  typing.  Also, a new near-IR spectrum obtained with the IRTF/SpeX
  spectrograph confirms an integrated-light spectral type of T5
  (\S~\ref{sec:spex}).}  This object was first resolved as a
0.065\arcsec\ binary in August~2000 in \HST/WFPC2 imaging
\citep{2003ApJ...586..512B}.  Among known visual ultracool binaries in
the field, this system has the shortest estimated orbital period, only
4~years (see compilation in \citealp{2006astro.ph..2122B}).  In
combination with its very high quality parallax measurement of
$73.6\pm1.2$~mas \citep{2003AJ....126..975T} and the fact that its two
components are nearly equal-magnitude (indicating nearly equal
masses), this system is a very appealing object for astrometric
monitoring.

Laser guide star (LGS) AO provides a powerful tool for high angular
resolution studies of brown dwarf binaries.  Through resonant
scattering off the sodium layer at $\sim$90~km altitude in the Earth's
atmosphere, sodium LGS systems create an artificial star bright enough
to serve as a wavefront reference for AO correction
\citep{1985A&A...152L..29F, 1987Natur.328..229T, 1994OSAJ...11..263H}.
Thus, most of the sky can be made accessible to near
diffraction-limited IR imaging from the largest existing ground-based
telescopes.  We have previously used Keck LGS AO to discover that the
nearby L~dwarf Kelu-1 is a binary system \citep{2005astro.ph..8082L}
and to identify the novel L+T binary SDSS~J1534+1615AB
\citep{2006astro.ph..5037L}.
In regards to dynamical mass determinations, ground-based telescopes
equipped with LGS AO can provide the necessary long-term platforms for
synoptic monitoring of visual binaries, especially where the required
amount of observing time at each epoch is relatively modest but many
epochs are needed, in contrast to \HST\ where target acquistion can be
slow and monitoring a populous sample over many epochs is quite
telescope time-intensive.

We present here the results of multi-epoch imaging of \twomassbin,
observed as part of our ongoing high angular resolution study of
ultracool binaries using LGS AO.  \S~2 presents our Keck LGS AO
observations and (re-)analysis of archival \HST\ imaging.  \S~3
presents the resolved photometric properties of the binary and fitting
of the orbit using a Markov Chain Monte Carlo method.  \S~4 compares
the resulting total mass against evolutionary models, and \S~5
summarizes our findings.  Those readers interested solely in the
results can focus on \S~4 and~\S~5.

\section{Observations}

\subsection{Keck LGS AO}

We imaged \twomassbin\ from 2005--2008 using the sodium LGS AO system
of the 10-meter Keck II Telescope on Mauna Kea, Hawaii
\citep{2006PASP..118..297W, 2006PASP..118..310V}.  Conditions were
photometric for all the runs.  We used the facility IR camera NIRC2
with its narrow field-of-view camera, which produces a $10.2\arcsec
\times 10.2\arcsec$ field of view.  Setup times for the telescope to
slew to the science targets and for the LGS AO system to be fully
operational ranged from 7--20~min, with an average of 12~min
\citep[e.g.][]{2006SPIE.6272E..14L}.

The LGS provided the wavefront reference source for AO correction,
with the exception of tip-tilt motion.  The LGS brightness, as
measured by the flux incident on the AO wavefront sensor, was
equivalent to a $V\approx9.2-10.3$~mag star.  Tip-tilt aberrations and
quasi-static changes in the image of the LGS as seen by the wavefront
sensor were measured contemporaneously with a second, lower-bandwidth
wavefront sensor monitoring the $R=16.2$~mag field star
USNO-B1.0~0601-0344964 \citep{2003AJ....125..984M}, located 31\arcsec\
away from \twomassbin.

At each epoch, \twomassbin\ was imaged in filters covering the
standard 2.2~\micron\ atmospheric window from the the Mauna Kea
Observatories (MKO) filter consortium \citep{mkofilters1,
  mkofilters2}.  Our initial observations in April~2005 were carried
out with the $\Kp$ (2.12~\micron) filter to minimize the thermal
background from the AO system, which is kept at ambient temperature.
Subsequent runs employed the $K$ (2.20~\micron) or $\Ks$
(2.15~\micron) filters.  Hereinafter, for brevity we refer to all
these data simply as $K$-band observations.

On each observing run, we typically obtained a series of dithered
$K$-band images, offsetting the telescope by a few arcseconds between
each 1--2 images.  The sodium laser beam was pointed at the center of
the NIRC2 field-of-view for all observations.  In April~2005, we also
obtained images with the MKO $J$ (1.25~\micron) and $H$ (1.64~\micron)
filters.  In April 2008, we also obtained images with the $CH_4s$
filter, which has a central wavelength of 1.592~\micron\ and a width
of 0.126~\micron; this filter is positioned around the $H$-band flux
peak in the spectra of mid/late-T dwarfs.

The images were reduced in a standard fashion.  We constructed flat
fields from the differences of images of the telescope dome interior
with and without continuum lamp illumination.  Then we created a
master sky frame from the median average of the bias-subtracted,
flat-fielded images and subtracted it from the individual images.
Images were registered and stacked to form a final mosaic, though all
the results described here were based on analysis of the individual
images.  Outlier images with much poorer FWHM and/or Strehl ratios
were excluded from the analysis.  Instrumental optical distortion was
corrected based on analysis by B. Cameron (priv.\ comm.) of images of
a precisely machined pinhole grid located at the first focal plane of
NIRC2.  The 1$\sigma$ residuals of the pinhole images after applying
this distortion correction are at the 0.6~mas level over the detector
field of view.  Since the binary separation and the imaging dither
steps are small, the effect of the distortion correction is minor,
smaller than our final measurement errors.

\begin{figure}
\centerline{\includegraphics[width=5.5in,angle=0]{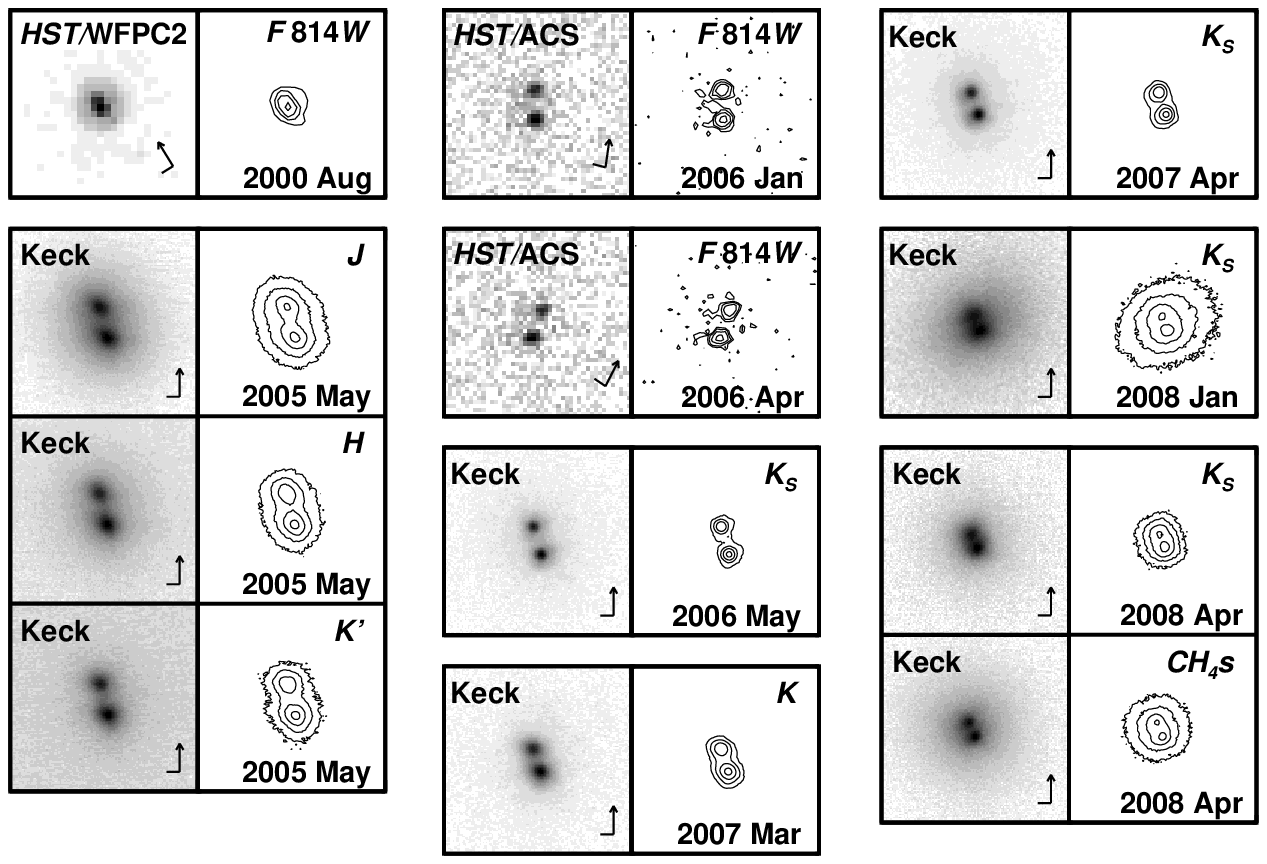}}
\caption{\normalsize Images of \twomassbin\ from \HST/WFPC2, Keck LGS,
  and \HST/ACS, arranged chronologically in each column.  Each image
  is 1.24\arcsec\ (16.8~AU) on a side, with the orientation indicated
  by the compass roses.  Note that the sky directions for the ACS
  images are not orthogonal, due to optical distortion in this
  instrument.  (We chose not to rotate the \HST\ images to the
  orientation of the Keck image for this figure, but instead to
  display the actual images so as to preserve the quality of the
  data.)  The greyscale images use a square-root stretch.  The
  contours are drawn from 80, 40, 20 and 10\% of the peak pixel.  Due
  to the effect of atmospheric dispersion, the Keck LGS $J$ and
  $H$-band images are slightly elongated in the vertical direction
  (which coincided with the elevation axis for these
  observations). \label{fig:images}}
\end{figure}

Table~\ref{table:keck} compiles the details of our observations, and
Figure~\ref{fig:images} presents our Keck LGS data.  Full widths at
half maxima (FWHM) and Strehl ratios were determined from two field
stars located $\approx$5--6\arcsec\ from \twomassbin.  The tabulated
errors on the FWHM and Strehl ratios are the standard deviation of
these quantities as measured from the individual images.

To measure the flux ratios and relative positions of \twomassbin's two
components, we mostly used the two aforementioned nearby field stars,
observed simultaneously with \twomassbin\ on NIRC2.  These stars
provided an excellent measurement of the instantaneous PSF.  We
empirically modeled the PSF using the Starfinder software package
\citep{2000A&AS..147..335D}, which is designed for analysis of blended
AO images.  For the Jan~2008 data, we employed a different procedure,
fitting analytic PSFs comprising multiple elliptical gaussians to
model the binary images.  These data were taken at much higher airmass
than all the other data.  Because of the larger atmospheric dispersion
and the different colors of the field stars relative to
\twomassbin, PSF fitting produced less accurate results than the
analytic approach, as determined by the artificial binary tests
described below.  
For every image, we fitted for the fluxes and positions of the two
components and then computed the flux ratio, separation, and position
angle (PA) of the binary.  The averages of the results were adopted as
the final measurements.  Overall, our PSF fitting produced very high
quality relative measurements, with errors of order 1\% for the flux
ratios, 0.05~pixels for the binary separation, and 0.2\degs\ for the
PA.  Note that these latter two values account only for the internal
instrumental measurements and do not include the errors on the
astrometric calibration of NIRC2, which we include below.

In order to gauge the accuracy of our measurements, we created myriad
artificial binary stars from images of the two PSF stars.  One PSF
star was used to create artificial binaries, and the other was used as
the single PSF for fitting the components.  For data at each epoch,
Starfinder was applied to the artificial binaries with similar
separations and flux ratios as \twomassbin.  These simulations showed
that any systematic offsets in our fitting code are very small, well
below the random errors, and that the random errors are accurate.  In
cases where the RMS measurement errors from the artificial binaries
were larger than those from the \twomassbin\ measurements, we
conservatively adopted the larger errors.\footnote{We also
  experimented with directly fitting the binary data by itself,
  without using other stars as the PSF.  This approach was similar to
  our previous analyses \citep{2005astro.ph..8082L,
    2006astro.ph..5037L}, namely we fit the images of the binary with
  either (1) an analytic model of the point spread function (PSF) as
  the sum of elliptical Gaussians, or (2) an empirical model derived
  iteratively using Starfinder-based code.  These were also tested
  against images of simulated binaries.  As before, we found that the
  Starfinder measurements were somewhat better compared to the
  analytical fits for the $K$-band data, and the multi-Gaussian
  fitting was better for the $J$ and $H$-band data.  Overall, fitting
  the binary images by themselves produced very good astrometry ---
  with internal errors (\ie, without the uncertainty in the absolute
  NIRC2 astrometric calibration) of about 3\% for the flux ratio,
  0.1~pixels for the separation, and 0.3\degs\ for the PA --- but
  slightly worse than our fits using field stars as PSFs.  The
  exception was the Jan~2008 data, as described in the text.}

To convert the instrumental measurements of the binary separation and
PA into celestial units, we used a weighted average of the calibration
from \citet{2006ApJ...649..389P}, with a pixel scale of
$9.963\pm0.011$~mas/pixel and an orientation for the detector's
+y~axis of $-0.13\pm0.07\degs$ east of north.  These values agree well
with Keck Observatory's notional calibration of
$9.942\pm0.05$~mas/pixel and $0.0\pm0.5$\degs, as well as the
$9.961\pm0.007$~mas/pixel and $-0.015\pm0.134$\degs\ reported by
\citet{2007AJ....133.2008K}.  Also, comparison of NIRC2 images of M92
to astrometrically calibrated \HST/ACS Wide-Field Camera images gives
a pixel scale for NIRC2 that agrees to better than 1~part in 10$^{-3}$
with our values (J. Anderson, priv.\ comm.).

Finally, we must consider the effect of atmospheric refraction.
Because of the southern declination of \twomassbin, all of our Keck
observations were necessarily undertaken at significant airmass
($>$1.55).  Because the two components of the binary do not have
exactly the same spectral types (\S~\ref{sec:phot}), the observed
positions on the sky are subject to slightly different amounts of
differential chromatic refraction (DCR).  We computed the expected
shift in the relative astrometry at each epoch using the prescriptions
of \citet{1992AJ....103..638M} for the DCR offset and
\citet{1984A&A...138..275S} for the refractive index of dry air.  We
assumed a fiducial temperature of 275 K and pressure of 608 millibars
for conditions on Mauna Kea \citep{1988PASP..100.1582C}.  We computed
the effective wavelengths for spectral types of T5.0 and T5.5 for the
two components using using all available spectra of these subclasses
contained in the SpeX Prism Spectral Library (from
\citealp{2004AJ....127.2856B, chiu05}, and
\citealp{2007AJ....134.1162L}) and the appropriate filter response
curve.\footnote{MKO filter curves available at
  ftp://ftp.jach.hawaii.edu/pub/ukirt/\~{}skl/filters.}  Note that
because of the unusual spectra of T~dwarfs, the effective wavelengths
of the secondary component is actually bluer in $H$- and $K$-bands and
redder only in $J$-band compared to the primary component.  Given the
fact that the secondary is mostly north of the primary at all of our
Keck epochs, DCR causes the separation of the binary to appear
slightly smaller at $H$ and $K$-bands and slightly larger at $J$-band
compared to the true position as would be observed at zenith.
The amplitude of the DCR effect is about 0.3~mas, much smaller than
the measurement errors at most (but not all) of the Keck epochs.
However, the effect is a systematic one so we correct the relative
astrometry of the two components based on our calculations.

Table~1 presents the final resulting measurements from our Keck LGS
data.  For the April~2005 dataset, all three filters give astrometry
consistent within the measurement errors; we use only the $H$-band
results in the orbit fitting discussed below.  In the Table and in our
orbit fitting (\S~\ref{sec:orbit-fitting}), we take care to
discriminate between the instrumental errors (namely those that arise
solely from fitting the binary images) and the overall astrometric
calibration of NIRC2, and thus any future refinements in the latter
can be readily applied to our measurements.

\subsection{\HST}

\subsubsection{WFPC2 Planetary Camera \label{sec:wfpc2}}

The two components of \twomassbin\ are only barely resolved in the
\HST/WFPC2 $F814W$ discovery images from August~2000.  Therefore, to
determine their relative positions and fluxes, we must model the
images using the sum of two blended PSFs.  The PSF of WFPC2's
Planetary Camera (PC) is undersampled (FWHM~=~1.7~pix for $F814W$);
this makes any empirical determination of the PSF difficult without
PSFs sampled at many subpixel locations.  Moreover, Anderson \& King
(2003) found that the WFPC2 PSF varies significantly over the detector
due to geometric distortion, making it impossible to construct a
reliable empirical PSF from other stars in the same image, even if
there are enough to sample many subpixel locations.  The original
analysis by \citet{2003ApJ...586..512B} employed a hybrid
gaussian/empirical PSF to fit for the binary parameters with resulting
uncertainties of $\pm$7~mas in separation and $\pm$9$^{\circ}$ in PA.
The astrometry from the WFPC2 discovery epoch is obviously very
important to the orbit determination, so we undertook our own analysis
with a more accurate PSF model to improve the precision of the binary
parameters.  

We used the TinyTim software package (Krist 1995) to create model PSFs
for the WFPC2 images.  We generated 5$\times$ super-sampled PSFs that
included the effects of (1) variation with position on the detector;
(2) broad-band wavelength dependence, by taking into account the
filter response curve and the spectrum of the source (using the
Keck/LRIS optical spectrum of the T4.5 dwarf 2MASS~J05591914$-$1404488
from Burgasser et al. (2003) as the template for the individual
components of \twomassbin), (3) telescope jitter (0 to 20~mas of
gaussian jitter), and (4) telescope defocus ($\pm$10 $\mu$m) to
account for \HST\ breathing effects. Because the geometrical
distortion is location-dependent, we used TinyTim model PSFs generated
for the nearest integer pixel location to the centroid of the binary
or single T~dwarf.  Also, we used the template spectrum closest to the
spectral type of the T dwarf with sufficient wavelength coverage
(0.70--0.96 $\mu$m) from S. K. Leggett's spectral
library.\footnote{The optical spectra we used were:
  2MASS~J05591914$-$1404488 for T4.5 and T5.5 objects
  \citep{2003ApJ...594..510B}; SDSSp~J162414.37$+$002915.6 for T6
  objects \citep{2000AJ....120.1100B}; SDSSp~J134646.45$-$003150.4 for
  T6.5 objects \citep{2000AJ....120.1100B}; and Gl~570D for T7.5
  objects \citep{2003ApJ...594..510B}.}

These TinyTim model PSFs were used to fit simultaneously for (1) the
location of the primary, (2) the location of the secondary, (3) the
normalization of the model PSF to the primary, and (4) the flux ratio
of the two components.  When fitting positions, the super-sampled
TinyTim PSF was interpolated using cubic convolution to the
appropriate subpixel location.  The best fit values were found using
the \texttt{amoeba} alogorithm \citep[e.g.,][]{1992nrca.book.....P} to
find the minimum $\chi^2$ value of a 1.1''$\times$1.1'' subimage
centered on the binary.  The image was cleaned using the IDL routine
\texttt{CR\_REJECT} in the Goddard IDL library to identify and mask
the numerous cosmic rays in the undithered WFPC2 image pair.  Masked
pixels were excluded from the computation of the $\chi^2$ value.  The
noise in each pixel was determined from the bias-subtracted raw WFPC2
image, assuming a read noise of 5.3 e$^-$/pix and a gain given by the
header keyword \texttt{ATODGAIN}.  A grid of PSFs in telescope jitter
and defocus were tried, and the fit corresponding to the jitter and
defocus combination yielding the lowest $\chi^2$ was chosen.  For
images of \twomassbin\, we found that our PSF-fitting routine yielded
residuals of $\lesssim$2\% of the peak value in 90\% of pixels with
$S/N > 3$.

Due to the optical distortion and ``34-th row'' defect present in the
WFPC2 (Anderson \& King 1999, 2003), the best-fit pixel locations of
each binary component do not exactly correspond to their locations in
an undistorted celestial reference frame.  To remove these effects, we
applied the corrections of Anderson \& King (2003), using a pixel scale
of $45.54\pm0.01$~mas/pix.\footnote{The value of the pixel scale and
  its uncertainty come from the WFPC2 Instrument Handbook for Cycle
  13.  This number is consistent with other measurements available in
  the literature: (1)~Holtzman et al. (1995) derived a pixel scale of
  45.54 mas/pix by comparing commanded telescope offsets in arcsec to
  the resulting pixel offsets; (2)~Holtzman et al. also derived a
  pixel scale of 45.55 mas/pix by comparison to an astrometric standard
  field in M67; (3) Pascu et al. (1998) used the JPL ephemeris of the
  satellites of Uranus to derive a pixel scale of 45.57 mas/pix; and
  (4) Pascu et al. (2004) used the JPL ephemeris of the satellites of
  Neptune to derive a pixel scale of 45.55~mas/pix.  The scatter in
  these pixel scales is consistent with our quoted uncertainty.}
The 34-th row effect could change, for example, the binary separation
by as much as 0.7~mas (a systematic shift of $-$0.6$\sigma$), but
because the binary components do not straddle a defective row and are
separated by a mere 1.4~pixels, the application of the 34-th row and
distortion corrections have a negligible effect on the astrometry.

With only two undithered WFPC2 images of \twomassbin, it is
challenging to quantify the measurement uncertainties, and it is
impossible to completely quantify the systematic errors, which arise
from an imperfect PSF model and also probably depend on the subpixel
positions of the two components given the undersampled nature of the
data. Using only the RMS scatter of the two measurements, the inferred
random errors in separation, PA, and flux ratio are 0.9 mas,
0.07$^{\circ}$, and 0.03 mag, respectively. To derive more robust
random errors and to investigate the systematic errors, we conducted
an extensive Monte Carlo simulation of our fitting routine.

We used WFPC2 $F814W$ images of seven other T~dwarfs from the same
\HST\ program, all apparently single, to create artificial binaries
which we then modeled using our PSF-fitting routine.\footnote{These
  were 2MASS~J05591914$-$1404488 (T4.5), 2MASSI~J0937347$+$293142
  (T6.0), 2MASSI~J1217110$-$031113 (T7.5), 2MASS~J12373919$+$6526148
  (T6.5), Gl~570D (T7.5), 2MASSI~J1546291$-$332511 (T5.5), and
  2MASSI~J2356547$-$155310 (T5.5), using spectral types from
  \citet{2005astro.ph.10090B}. The remaining object from this program,
  2MASSI~J1047538$+$212423 (T6.5), was unusable for our purposes
  because it landed on a bad column.} Because the WFPC2 PSF is
severely undersampled, we only created artificial binaries with
integer-shifted positions. It turns out that the WFPC2 locations of
the two components of \twomassbin\ are at a very nearly
integer-shifted separation of 1.4~pixels ($\Delta{x}\approx1$ pix,
$\Delta{y}\approx-1$ pix).  Therefore, in determining the
uncertainties and systematic offsets we used only the configuration
most nearly matching that of \twomassbin\, with a separation of
$\sqrt{2}$ pix and instrumental PA of 225$^{\circ}$. We found that
using any or all of the other three $\sqrt{2}$ configurations gave
consistent uncertainties.
After subtracting the best-fit model, artificial binary images yielded
residual images in which 90\% of pixels with $S/N > 3$ were below
1--3\% of the peak flux, comparable to the residual images of
\twomassbin.

The images of the single T~dwarfs span a range in $S/N$ from about
1.5~mag brighter to 1.3~mag fainter than the primary component of
\twomassbin.  We used these images at their native $S/N$ when
simulating the primary component.  To simulate the secondary
component, we degraded the $S/N$ of the images assuming a flux ratio
of 0.30~mag.  We also tried flux ratios of 0.25~mag and 0.35~mag to
explore the possibility that the uncertainties depend on the assumed
flux ratio, but we found that this had an insignificant effect on our
predicted uncertainties ($<1\sigma_{\sigma}$, where $\sigma_{\sigma} =
\sigma / \sqrt{2N_{sim}}$ and $N_{sim}$ is the number of simulations).
Signal-to-noise degradation of an image was done by a multiplicative
scaling followed by the addition of normally distributed random noise
to each pixel, according to the same WFPC2 noise model we used to
determine $\chi^2$ in the PSF-fitting procedure. In fact, by running
simulations where the primary images were degraded to much lower
$S/N$, we found that all of the single T dwarfs are in a high $S/N$
regime in which systematic errors (PSF model imperfections) dominate,
not random errors (photon noise) --- our simulations showed no
dependence between the $S/N$ of the T~dwarf used to construct
artificial binaries and the resulting astrometric uncertainties.
Therefore, we used the RMS of the results from all simulated binaries
in order to determine the final uncertainties for \twomassbin.  As
expected, the uncertainties in separation, PA, and flux ratio from our
simulations were somewhat larger than those derived from the standard
deviation of the measurements from the two \twomassbin\ images, since
both random and systematic errors have been evaluated in the
simulations.  In fact, the larger uncertainties are not simply due to
averaging over, for example, the many subpixel locations of the single
T~dwarfs used in the Monte Carlo, because the simulated measurements
for each single T~dwarf show scatter consistent with the final derived
uncertainties.

Table~\ref{table:hst} presents our final results for the WFPC2 images,
with systematic offsets from the Monte Carlo simulations applied.  Our
astrometry agrees well (better than 1$\sigma$) with the original
results of \citet{2003ApJ...586..512B}, though our measurement errors
are a factor of~8 smaller.  Part of this improvement comes from our
use of TinyTim-computed PSFs, as opposed to the simpler Burgasser
\etal\ PSF model of a gaussian plus empirical residuals.  We also used
all possible single PSFs in our artificial-binary simulations, whereas
Burgasser \etal\ used only WFPC2 images of 2MASS~J0559$-$1404, a
source that is suspected to be an unresolved
binary.\footnote{2MASS~J0559$-$1404 is roughly twice as luminous as
  objects of similar spectral type \citep{2002AJ....124.1170D,
    2003AJ....126..975T, 2004AJ....127.2948V}.  High angular
  resolution observations have not detected any multiplicity
  (\citealp{2003ApJ...586..512B}; Liu \etal, in prep.), but the source
  could be a very tight system, \eg, with a 0.5~pixel separation.  If
  so, its multiplicity could confuse any attempts to fit only two
  single PSFs to the tightest artificial binaries constructed from its
  image.  Interestingly, we found that at the smallest (1.0 pixel)
  separations, artificial binaries constructed from images of
  2MASS~J0559$-$1404 yielded extremely large uncertainties (5~mas and
  15\degs). We did not observe such behavior for any of the other six
  apparently single T dwarfs, nor were the uncertainties for larger
  separation binaries made from 2MASS~J0559$-$1404 abnormally large.
  One natural explanation would be that the source is just marginally
  resolved in the WFPC2 imaging.}
And part of the improved uncertainties is somewhat illusional, as it
arises from the different parameter space explored in Monte Carlo
simulations by us and Burgaser \etal\ As a check, we ran a suite of
simulations more comparable to that of Burgasser \etal, in which the
ranges of artificial binary parameters were 1.0--3.0 pixels in
separation, the full range of PAs, and 0.0--1.0 mag in flux ratio.
These yielded similar astrometric uncertainties to the published
errors, suggesting that the smaller uncertainties we derive are due to
our more restricted choice of artificial binary configurations
($\Delta{x}=1$ pix, $\Delta{y}=-1$ pix) and/or averaging over many
single PSFs to reduce systematics associated with any one specific
object.

Our improvement to the WFPC2 astrometry was essential in our early
attempts to fit the orbit based on Keck data obtained in 2005--2007.
However, with the addition of data in 2008, the final \HST+Keck
dataset has sufficent time baseline and astrometric quality that the
choice of WFPC2 astrometry does not highly impact the final orbit
fitting results (\S~\ref{sec:altorbits}).

\subsubsection{ACS High Resolution Camera}

The \twomassbin\ system was observed on 2006~January~19 and
2006~April~11 (UT) with the High Resolution Camera (HRC) of \HST's
Advanced Camera for Surveys (ACS) by program GO-10559 (PI H. Bouy).
The binary is much more widely separated at these epochs than in the
WFPC2 observations, but the PSFs of the two components are still
blended.  We have therefore applied the same TinyTim PSF-fitting
technique described in the previous section to derive the relative
astrometry from the ACS images.  The primary differences between the
WFPC2 and ACS datasets are: (1) ACS has much more severe geometric
distortion than WFPC2, which changes the shape of the PSF and
complicates astrometry because the pixels projected on the sky are not
square, and (2) the ACS data are of much lower $S/N$, with a total
exposure time of only 50~sec for each cosmic-ray rejected, combined
dithered image (\cf, 1300~sec for a single WFPC2 image).  Because of
the lower $S/N$, we found it unwarranted to fit the ACS images for
telescope jitter and defocus as adding these free parameters did not
improve the quality of the fits (as verified in the Monte Carlo
simulations discussed below).  Also, we found that the
$\approx$25$\times$ lower $S/N$ of these data almost exactly negates
any improvement to the astrometry that might be expected given the
larger binary separation at these epochs.

We used distorted model PSFs generated by TinyTim to fit for the
position and flux of each binary component in images that had been
cosmic-ray cleaned (\texttt{CRSPLIT=4}) by the latest \HST\ pipeline.
Best-fit pixel locations were corrected for geometric distortion using
the solution of Anderson \& King (2004; Instrument Science Report
04-15), and we used their measured ACS pixel scale, which was derived
by comparing commanded (\texttt{POSTARG}) offsets of \HST\ in
arcseconds to the resulting pixel offsets.  They derived two such
pixel scales for two epochs of observations of 47~Tuc, and we adopt the
mean and standard deviation of these two:
$28.273\pm0.006$~mas/pix.\footnote{Comparison of WFPC2-PC images of
  47~Tuc (GO-8267, PI Gilliland) with ACS-WFC images (GO-10775, PI
  A. Sarajedini) shows excellent agreement between our adopted pixel scales
  for the two instruments, at the level of $2\times10^{-4}$ which is
  well below the other errors in the measurements (J. Anderson, priv.\
  comm.).}  For each epoch, we adopted the mean of the measurements
from all four dithered images for the binary parameters of \twomassbin\
(Table~\ref{table:hst}).

Again, to investigate the measurement errors thoroughly, we performed
Monte Carlo simulations of our fitting routine.  We used images of
single brown dwarfs to construct artificial binaries in configurations
resembling \twomassbin.  At both ACS epochs, the binary is well
represented by integer-pixel shifts on a grid where $\Delta{x}=(6,7)$
and $\Delta{y}=(-3, -2, -1, 0, 1)$.  There have been no \HST/ACS
science programs dedicated to studying single brown dwarfs; however,
ACS images of the single brown dwarfs 2MASS~J00361617$+$1821104 (L3.5;
\citealp{2000AJ....120..447K}) and 2MASS~J05591914$-$1404488 (T4.5;
\citealp{2005astro.ph.10090B}) were obtained for calibration purposes
and are available in the \HST\ Archive (CAL/ACS-10374, PI Giavalisco).
We found that despite the large difference in spectral types, any
corresponding difference in the PSF does not alter the results of the
Monte Carlo simulations. The $S/N$ of each of these single objects is
much higher than that of \twomassbin, so we degraded the $S/N$ of the
single brown dwarfs for the artificial binary simulations. In fact, by
varying the $S/N$ of the simulations, we found that the ACS data for
\twomassbin\ are well in the $S/N$ regime dominated by random photon
noise, while the images of the single objects are in a high $S/N$
regime dominated by systematic errors (akin to the WFPC2 images of
\twomassbin).  Therefore, given that we have four images, we divide
the RMS of the Monte Carlo results by $\sqrt{4}$ to represent the
final uncertainties.

Table~\ref{table:hst} contains our final ACS results.\footnote{After
  our paper was submitted, \citet{2008A&A...481..757B} reported an
  analysis of the same ACS images.  Their results agree with ours to
  within the stated uncertainties.  Their errors are slightly smaller
  than ours in separation (0.9~mas compared to 1.1~mas) and much
  smaller in PA (0.1\degs\ compared to 0.5\degs).}  Note that the ACS
data are contemporaneous with our Keck LGS data, and observations by
the two telescopes in 2006 separated by less than one month show
excellent agreement.  However, the relatively low $S/N$ of the ACS
data means it has the larger astrometric errors.  Our simulations
confirm that if the ACS data were of higher $S/N$ (i.e., longer
exposure times than 50 s), the resulting astrometric precision would
be much better than, instead of comparable to, the WFPC2 astrometric
uncertainties.


\subsection{IRTF/SpeX Spectroscopy \label{sec:spex}}

\begin{figure}
\centerline{\includegraphics[height=7.5in,angle=0]{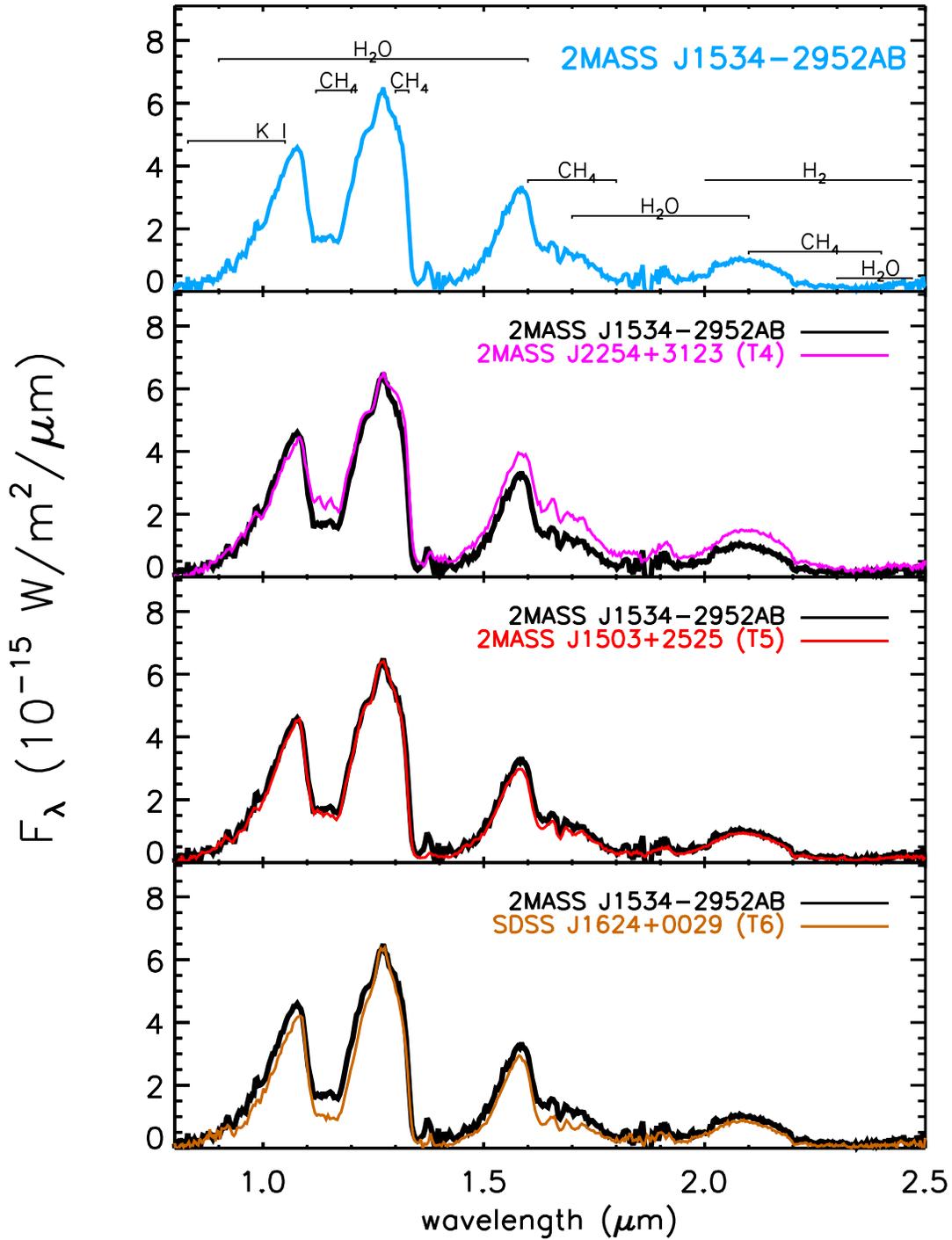}}
\caption{\normalsize {\bf Top panel:} Near-IR spectrum of \twomassbin\
  obtained with the IRTF/SpeX spectrograph.  {\bf Other panels:} Same
  spectrum of \twomassbin\ plotted as a thick black line.  Spectra of
  T~dwarf spectral standards from \citet{2005astro.ph.10090B} are
  plotted as colored lines.  The spectra have been normalized by their
  peak flux.  \label{fig:spectra}}
\end{figure}

We obtained low-resolution ($R\approx$150) integrated-light spectra of
\twomassbin\ on 2008~May~16~UT from NASA's Infrared Telescope Facility
(IRTF) located on Mauna Kea, Hawaii.  Conditions were photometric with
seeing of about 0.7\arcsec\ FWHM near the target.  We used the
facility near-IR spectrograph Spex \citep{1998SPIE.3354..468R} in
prism mode, obtaining 0.8--2.5~\micron\ spectra in a single order.  We
used the 0.5\arcsec\ wide slit, oriented at the parallactic angle to
minimize the effect of atmospheric dispersion.  \twomassbin\ was
nodded along the slit in an ABBA pattern, with individual exposure
times of 180~sec, and observed over an airmass range of 1.64--1.60 as
it rose.  The telescope was guided during the exposures using images
obtained with the near-IR slit-viewing camera.  The total on-source
exposure time was 720~sec.  We observed the A0~V star HD~142851
contemporaneously for flux and telluric calibration.  All spectra were
reduced using version 3.4 of the SpeXtool software package
\citep{2003PASP..115..389V,2004PASP..116..362C}.  The reduced
IRTF/Spex spectrum is plotted in Figure~\ref{fig:spectra} and compared
to T~dwarf spectral standards from \citet{2005astro.ph.10090B}.
Visual examination shows an excellent match to the T5 spectral
standard 2MASS~J1503+2525, as does measurement of the
\citet{2005astro.ph.10090B} spectral indicies for \twomassbin:
\htwoo--$J$ = 0.271 (T4.8), \meth--$J$ = 0.420 (T4.8), \htwoo--$H$ =
0.345 (T5.0), \meth--$H$ = 0.430 (T5.0), and \meth--$K$ = 0.224
(T5.1), with spectral type estimates based on the polynomial fits to
the indices from \cite{burg2006-lt}.


\section{Results}

\subsection{Resolved Photometry and Spectral Types \label{sec:phot}}

We use our measured flux ratios and the published $JHK$ photometry
from \citet{2004AJ....127.3553K} to derive resolved IR colors and
magnitudes for \twomassbin\ on the MKO system.  We use the \HST\
photometry from \citet{2003ApJ...586..512B} in determining the resolved
$F814W$ magnitudes.  Then to infer spectral types for the individual
components, we compare these to magnitudes and colors of ultracool
dwarfs from \citet{2004AJ....127.3553K} and \citet{chiu05}, excluding
known binaries.
We use near-IR spectral classfications from the
\citet{2005astro.ph.10090B} scheme.
We assume that the components of \twomassbin\ are themselves single,
not unresolved binaries.  

\begin{figure}
\vskip -1in
\hskip 0.2in
\centerline{\includegraphics[width=6in,angle=90]{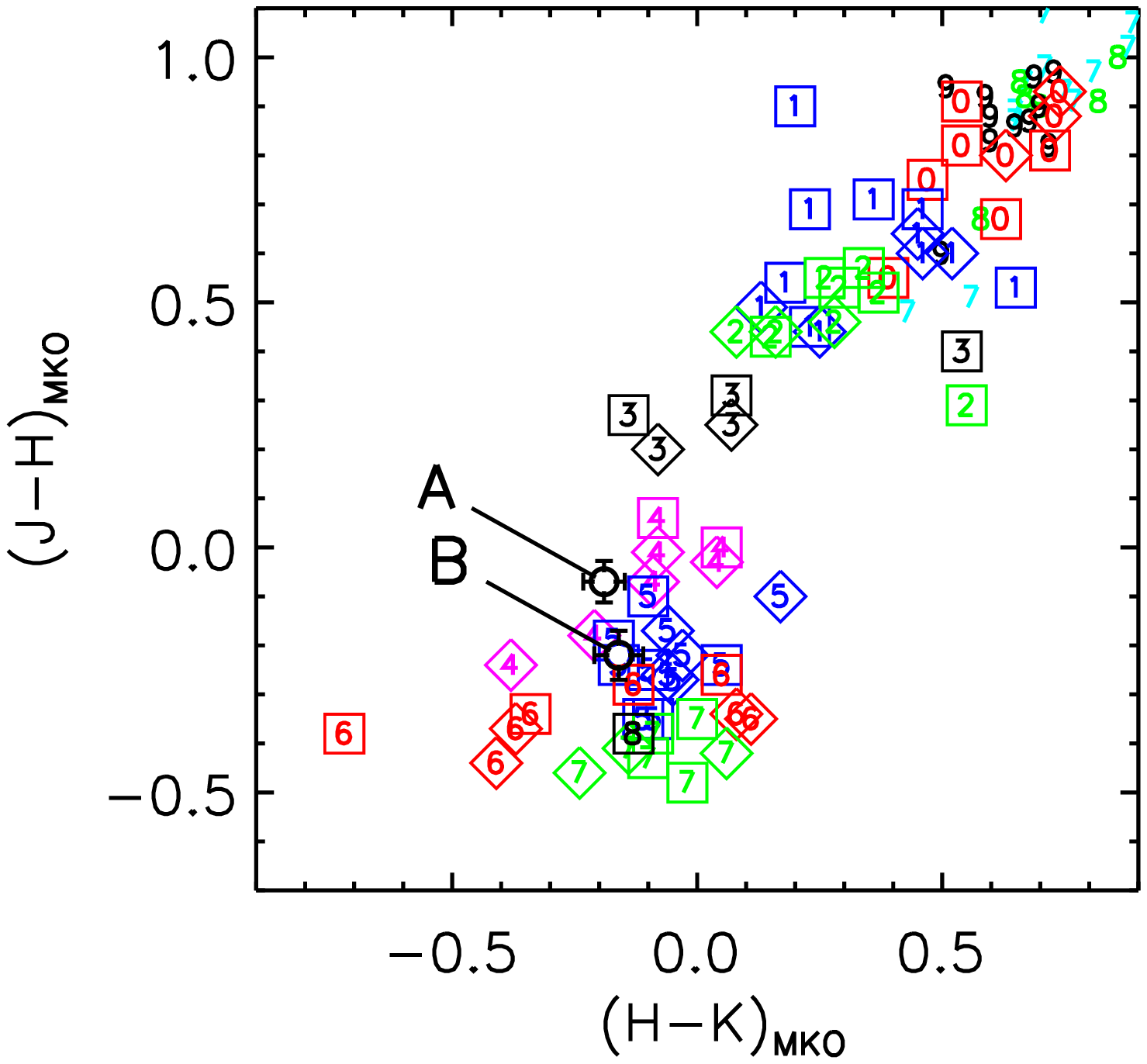}}
\vskip -2ex
\caption{\normalsize Near-IR colors of \twomassbin\ compared with
  nearby single late-L~and T~dwarfs from \citet{2004AJ....127.3553K}
  and \citet{chiu05} and individual components of resolved binaries
  from \citet{2004A&A...413.1029M, 2005astro.ph.10580B,
    2006ApJS..166..585B, 2005astro.ph..8082L, 2006astro.ph..5037L}.
  The photometry errors are comparable to or smaller than the size of
  the plotting symbols.  The numbers indicate the near-IR spectral
  subclass of the objects, with half subclasses being rounded down
  (\eg, T3.5 is labeled as ``3''), and objects of the same subclass
  plotted in the same color.  The late-L~dwarfs (classified on the
  \citealp{geb01} scheme) are plotted as bare numbers.  The T~dwarfs
  (on the \citealp{2005astro.ph.10090B} scheme) are plotted as
  circumscribed numbers, with squares for integer subclasses (\eg, T3)
  and diamonds for half subclasses (\eg, T3.5).  
  \label{fig:colorcolor}}
\end{figure}

Figure~\ref{fig:colorcolor} shows that component~A has IR colors most
typical of T4.5--T5 dwarfs, and component~B is most similar to T5--T6
dwarfs.  The individual absolute IR magnitudes (given in
Table~\ref{table:resolved}) give similar results.  The ``faint''
polynomial fits for absolute magnitude as a function of spectral type
from \citet{2006astro.ph..5037L} give $M(J$) = \{14.5, 14.6, 14.7,
14.9, 15.2\}~mag, $M(H)$ = \{14.6, 14.8, 15.0, 15.2, 15.5\}~mag, and
$M(K)$ = \{14.7, 14.9, 15.1, 15.4, 15.7\}~mag for near-IR spectral
types of T4.5, T5, T5.5, T6, and T6.5, respectively.
Averaging the same data for each individual subclass gives $M(J)$ =
\{13.9$\pm$0.6, 14.1, 14.4$\pm$0.4, 15.0$\pm$0.5, 15.1$\pm$0.5\}~mag,
$M(H)$ = \{14.0$\pm$0.6, 14.2, 14.6$\pm$0.4, 15.3$\pm$0.4,
15.4$\pm$0.5\}~mag, and $M(K)$ = \{14.0$\pm$0.5, 14.3,
14.6$\pm$0.4, 15.5$\pm$0.7, 15.6$\pm$0.9\}~mag, where the
uncertainties are the RMS of the photometry for each subclass (and no
listed uncertainties for subclasses with only one object).
Altogether, the absolute magnitudes suggest types of T5--T5.5 for
component~A and T5.5--T6 for component~B.

The resolved $(CH_4s-H)$ colors provides a third means to estimate the
spectral types, as these track the $H$-band methane absorption, which
correlates well with overall near-IR spectral type (\eg, Figure~2 of
\citealp{2005AJ....130.2326T}).  First, we compute individual $CH_4s$
magnitudes for \twomassbin\ using the $CH_4s$ flux ratio from our LGS
images, the integrated-light photometry of $H=14.74\pm0.03$~mag from
\citet{2004AJ....127.3553K}, and an integrated-light color of
$(CH_4s-H) = -0.33$~mag synthesized from the near-IR spectrum of
\citet{2005astro.ph.10090B}.  Including the measurement errors in the
flux ratios and $H$-band photometry, we find $CH_4s = 15.06\pm0.04$
and $15.27\pm0.04$~mag and $(CH_4s-H) = -0.30\pm0.05$ and
$-0.37\pm0.05$~mag for components A and B, respectively.  Note that
the {\em relative} $(CH_4s-H)$ color of the two components is known to
higher precision, since the above computed colors for the two
components contain the same 0.03~mag error that originates from the
integrated-light $H$-band photometry.  (In other words, the 0.05~mag
uncertainties in the colors of the two components are not independent,
but correlated.)  Removing this effect gives a relative color of
$\Delta(CH_4s-H) = 0.07\pm0.02$~mag between A~and~B, \ie, greater
methane absorption in component~B is detected.

\begin{figure}
\vskip -1in
\hskip 0.2in
\centerline{\includegraphics[width=5in,angle=90]{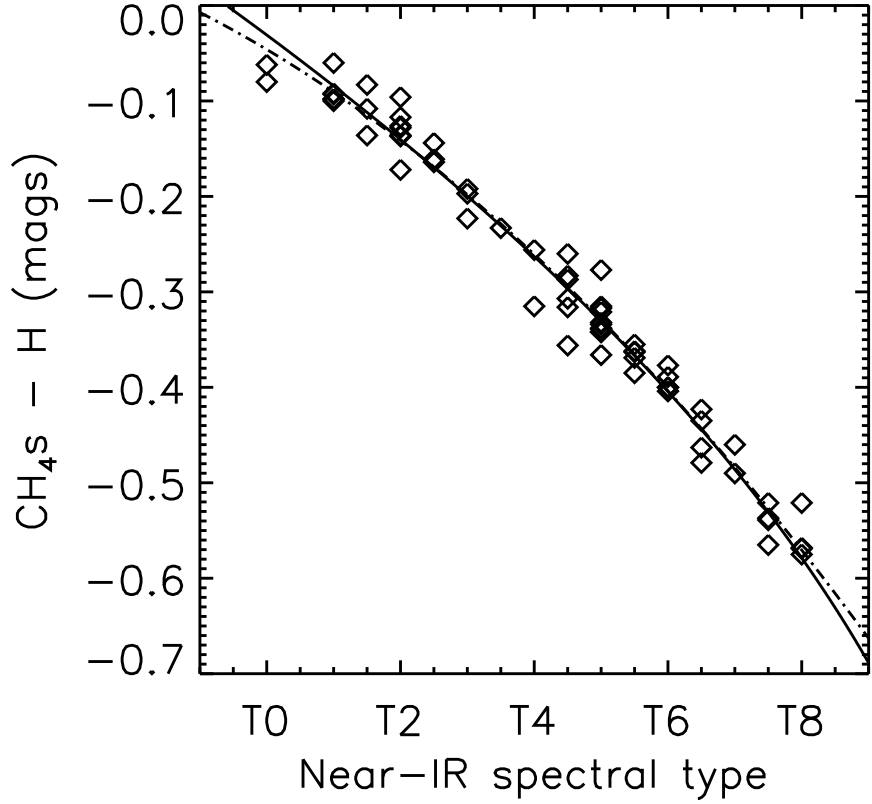}}
\vskip -2ex
\caption{\normalsize $(CH_4s-H)$ color versus near-IR spectral type
  for T~dwarfs.  The plotting symbols are the synthesized colors from
  published low-resolution near-IR spectra of T~dwarfs, excluding
  objects that are spectrally peculiar and known binaries.  (See
  \S~\ref{sec:phot} for references.)  The lines represent 2nd-order
  polynomial fits to the data, with the solid line being the spectral
  type as a function of color (\ie, the inverse fit) and the dotted
  line being the color as a function of spectral
  type.\label{fig:ch4short}}
\end{figure}

To determine the behavior of $(CH_4s-H)$ with near-IR spectral type,
we synthesized colors from the Spex Prism Spectral Library collection,
which contains low-resolution spectra of 68~T~dwarfs after removing
spectrally peculiar objects and known binaries.\footnote{{\tt
    http://www.browndwarfs.org/spexprism.}  The T~dwarf spectra are
  compiled from \citet{2004AJ....127.2856B, 2005astro.ph.10090B,
    2006ApJ...639.1095B, 2006astro.ph..9556B, chiu05,
    2007AJ....134.1162L, 2006astro.ph..9793L}.}
Figure~\ref{fig:ch4short} plots the results.  We fit a 2nd-order
polynomial for the dependence of color on near-IR spectral type and
vice-versa:
\begin{eqnarray}
(CH_4s - H)  & = & -0.400 + 0.07718 \times SpT - 0.0029736 \times SpT^2 \\
SpT          & = &  19.40 - 19.698 \times (CH_4s-H) - 8.3600 \times (CH_4s-H)^2
\end{eqnarray}
where $SpT=20$ for T0, =~21 for T1, etc.  The RMS scatter about the
fits are 0.02~mag and 0.3~subclasses, respectively.  Using these
polynomial relations, the observed $(CH_4s-H$) colors give spectral
types of T4.5~$\pm$~0.7 and T5.6~$\pm$~0.6 for components A and B,
respectively, where the spectral type uncertainties come from formal
propagation of the uncertainty in the colors.  In addition, just as
the relative $(CH_4s-H)$ color of the two components are known more
accurately than the absolute colors, we compute a relative spectral
type of $1.1\pm0.3$~subclasses between components~A and~B.

Combining the inferences from the $JHK$ colors, the absolute
magnitudes, and the $(CH_4s-H)$ colors, we adopt spectral type
estimates of T5~$\pm$~0.5 and T5.5~$\pm$~0.5 for the two components.
The relative $(CH4_s-H)$ color favors a slightly larger spectral type
difference than the absolute magnitudes but consistent with the
adopted uncertainty.  (Also the \Teff\ difference of the two
components computed in \S~\ref{sec:temp} favors a 0.5~subclass
difference.)

Higher-order multiple systems are very rare among ultracool binaries,
with an estimated frequency of $3_{-1}^{+4}\%$
\citep{2006astro.ph..2122B}, and thus a priori we do not expect
\twomassbin\ to fall into this category.  The colors and magnitudes
are consistent with the system being composed of only two components,
and not being a partially resolved higher order multiple system.  If
component~B was actually an equal-mass binary, the absolute magnitudes
of its components would be 0.75~mag fainter than the integrated-light
of~B, meaning $M(J,H,K) = \{15.5, 15.9, 16.0\}$~mags.  This would
suggest a spectral type around T7, based on the polynomial relations
in \citet{2006astro.ph..5037L}, which is clearly too late-type
compared to the integrated-light spectrum and the observed near-IR
colors of~B.


\subsection{Bolometric Luminosities \label{sec:lbol}}

To measure the \Lbol\ for the system, we combine our SpeX
0.9--2.4~\micron\ spectrum with the published integrated-light
$F814W$, \Lp-band, and \Spitzer/IRAC thermal-IR photometry and
uncertainties \citep{2003ApJ...586..512B, 2004AJ....127.3553K, gol04,
  2006ApJ...651..502P}.\footnote{The \Lp-band photometry for this
  system is anomalous, as \citet{2007ApJ...655.1079L} show the
  $([3.5]-\Lp)$ color is $\sim$0.2~mag redder than any other T~dwarf
  and $\sim$0.3~mag redder than objects of similar spectral type.
  Both of these bandpasses are affected by the 3.3~\micron\
  fundamental band of \meth, with the Spitzer [3.5] data being
  subjected to greater absorption, and thus the redder color might
  point to anomolously strong \meth.  However, the near-IR spectrum
  and the $JHK$ and Spitzer 3.5--7.9~\micron\ colors are consistent
  with other mid-T~dwarfs, and thus perhaps the \Lp-band photometry is
  incorrect.  Either way, this has a negligible effect on the computed
  \Lbol.}  We flux-calibrated the SpeX data using the published
$H$-band MKO photometry from Knapp \etal\ 2004.  For the \Spitzer/IRAC
data, we adopted the photometric calibration and an overall 2\%
absolute uncertainty based on Reach \etal\ (2005).  We extended the
binary's spectral energy distribution (SED) to short wavelengths by
linearly interpolating from the $F814W$ datum to zero flux at zero
wavelength and to long wavelengths by assuming a Rayleigh-Jeans
spectrum beyond the reddest Spitzer bandpass (7.87~\micron); this
extrapolation increases the total flux by 2\%.

We then integrated the SED, using a Monte Carlo approach to account
for all the measurement errors.  We find $\log(\Lbol/\Lsun) =
-4.751\pm0.011$~dex for the system, with the uncertainty increasing to
0.018~dex after including the uncertainty in the distance.  (As
discussed in \S~4, we keep track of these independent uncertainties in
our calculations.)  The largest uncertainty in the integration arises
from the 0.03~mag uncertainty in the integrated-light photometry used
to normalize the SpeX spectrum.  We cross-checked our method using the
same data for the T4.5 dwarf 2MASS~J0559$-$1404 and found excellent
agreement with the \Lbol\ measured by \citet{2006ApJ...648..614C}
using absolutely flux-calibrated spectra from 0.6--15~\micron.

The computed total \Lbol\ agrees well with that inferred from using
the $K$-band bolometric corrections ($BC_K$) from Golimowski \etal\
(2004), namely using the resolved $K$-band absolute magnitudes and the
estimated spectral types, which would give $\log(\Lbol/\Lsun) =
-4.97\pm0.06$ and $-5.06\pm0.06$~dex for the individual components and
thus $\log(\Lbol/\Lsun) = -4.71\pm0.08$~dex for the total system.
However, the uncertainties are larger when using $BC_K$ to derive
\Lbol, since this incorporates the uncertainties arising from the
0.5~subclass uncertainty (0.06~mag in bolometric magnitude) and the
intrinsic scatter in the Golimowski \etal\ $BC_K$ relation
(0.13~mag).  In short, direct integration of the observed SED is more
accurate.

To apportion the observed total \Lbol\ into the individual components,
we assume the observed $K$-band flux ratio of the system represents
the luminosity ratio.  This would be exactly correct if the two
components had identical spectral types (and neglecting photometric
variability).  To account for the difference in spectral types, we
generate a Monte Carlo distribution of $BC_K$ values for each
component using the Golimowski \etal\ polynomial fit as a function of
spectral type subject to the following rules: the spectral type of
component~A is uniformly distributed from T4.5--T5.5; the spectral
type of component~B is no later than T6; and the difference in their
spectral types is at least 0.5~subclasses.  This produces an average
difference in $BC_K$ between the two components of 0.09~mags and an
RMS of 0.03~mags.  Thus, we find $\log(\Lbol/\Lsun) = -5.015\pm0.019$
and $-5.093\pm0.019$~dex for the two components, including the
uncertainty in the distance.


\subsection{Dynamical Mass Determination \label{sec:orbit-fitting}}

\subsubsection{Orbit Fitting using Markov Chain Monte Carlo}

We have data at 9 independent epochs, which is formally
sufficient to determine the 7~parameters of a visual binary orbit
given our measurements (9 positions + 9 times).  However, two pairs of
measurements are separated by only one month (April/May 2006 and
March/April 2007), and the cadence of the orbital phase covered is
limited, with the \HST/WFPC2 datum being taken almost 5~years before
the next epoch.  While standard gradient-descent (Levenberg-Marquardt)
least-squares techniques would be sufficient to derive an orbit
(\S~\ref{sec:fitting}), we also would like to accurately determine the
probability distribution of the orbit parameters (which may not be
normally distributed) and the associated degeneracies.
For epochs with Keck data taken in multiple filters, we choose the
filter with the smallest astrometry errors.

Thus, we first used a combination of gradient-descent techniques from
random starting points and simulated-annealing techniques to isolate
the class of potential orbital solutions near a reduced chi-squared
(\rchisq) of~1.  Then, to fully explore this class of solutions, we
used a Markov Chain Monte Carlo (MCMC) approach
\cite[e.g.][]{bremaud99:markov_chain}.  MCMC provides a means to
explore the multi-dimensional parameter space inherent in fitting
visual orbits that is computationally efficient, able to discern the
degeneracies and non-gaussian uncertainites in the fit, and allows for
incorporation of a~priori knowledge.
In short, the MCMC approach is distinct from ordinary Monte Carlo
methods in that instead of a completely random steps through the model
parameter space, the steps are chosen such that the resulting number
of samples (the ``chain'') is asymptotically equivalent to the
posterior probability distribution of the parameters being sought.
(See \citealp{2004PhRvD..69j3501T}, \citealp{2005AJ....129.1706F}, and
\citealp{2005ApJ...631.1198G} for explications of applying MCMC to
astronomical data.)

We parameterized the binary's orbit using the standard 7 parameters:
period ($P$), semi-major axis ($a$), inclination ($i$), epoch of
periastron ($T_0$), PA of the ascending node ($\Omega$)\footnote{For
  visual binaries, there is a 180\degs\ ambiguity in determining
  $\Omega$, which can only be resolved with radial velocity
  information.}, eccentricity ($e$), and the argument of periastron
($\omega$).  We used an MCMC chain length of $2\times10^8$, with the
parameters stored every hundredth iteration.  We started the chain at
the global minimum found by the simulated-annealing and
gradient-descent algorithms.  We used the Metropolis-Hastings
algorithm to sample the joint probability distribution with a variant
of the usual Gibbs sampler.  Instead of chosing one of the
7~parameters to increment or decrement, we chose instead to move
randomly either forwards or backwards along one of 7~orthogonal
directions in the parameter space.  These directions were initially
chosen to be along the individual coordinate axes of our 7~parameters.
Then, every $5\times10^5$~iterations, the covariance matrix of the
7~parameters was calculated and new directions chosen along the
directions corresponding to the eigenvectors of the covariance
matrix. After covariance matrix calculation, the trial step size was
set to be the square root of the covariance matrix eigenvalues.  This
enabled the long, thin, curved minima in our parameter space to be
sampled much more efficiently.  Each set of $5\times10^5$ iterations
can thus be thought of as its own sub-chain with fixed directions. As
the first $5\times10^5$ iterations uses a less efficient sampling
(sampling directions aligned with the 7~parameters themselves), we
treated this part of the chain as the ``burn-in'' time, neglecting it
in the final analysis. As the chain was running, the trial step size
in each direction was continuously scaled over a timescale of
700~steps so that the success rate of jumps averaged to 0.25.  The
correlation length of our most correlated chain, as defined in
\citealp{2004PhRvD..69j3501T}, was $2\times10^4$ for the orbital
period ($P$), with equal or smaller correlation lengths for other
parameters.  This gives an effective length of the chain of
$\sim1\times10^4$, which in turn gives statistical uncertainties in the
parameter errors of about $1/\sqrt{10^4}=1$\%, \ie, negligible.  These
uncertainties in the errors are consistent with the results from
running multiple test chains.

By making steps of the same size in the positive and negative
directions for these parameters in constructing the Markov chain, we
would implicitly assume that our prior knowledge of these parameters
is a uniform distribution.  This is not an accurate representation of
our prior knowledge since, for example, binaries with periods between
$10^2$ and $10^3$ years are not 10~times more common than binaries
with periods between 10 and $10^2$ years.  We therefore applied a
prior to the likelihood function in the MCMC fitting where $P$ and $a$
are distributed evenly in logarithm and that the parameters
$e\cos(\omega)$ and $e\sin(\omega)$ are uniformly distributed, rather
than $e$ and $\omega$, to save the algorithm from unnaturally
preferring circular solutions.  This is equivalent to the $f(e)=2e$
distribution as discussed by, \eg, \citet{1991A&A...248..485D}. The
very small effect of the choice of prior is discussed
below.\footnote{Some care is also needed in handling the astrometric
  calibration errors of the three instruments used in our analysis.
  In practice, most of the constraints come from the \HST/WFPC2
  discovery epoch from August 2000 and the six Keck LGS epochs from
  2005--2008.  Given the large time difference, the two datasets
  essentially constrain different portions of the parameter space.
  Some extra information is provided by the \HST/ACS data but to a
  much lesser degree since it is contemporaneous with the Keck LGS
  data and has 2--4$\times$ larger errors.  (In fact, the MCMC fitting
  gives basically the same results if the ACS data are excluded.)
  Thus simply applying the NIRC2 astrometric calibration error of
  0.1\% in pixel scale and 0.07\degs\ in PA to all six Keck epochs
  would incorrectly treat this as a random error, when in fact the
  uncertainty in the NIRC2 calibration globally impacts the overall
  solution of, \eg, the semi-major axis and the PA of the line of
  nodes.  Therefore, we do the following: (1) we apply the errors in
  the NIRC2 calibration in quadrature to the \HST/ACS and \HST/WFPC2
  measurements before orbit fitting, and (2) once the semi-major axis
  distribution has been determined, we apply the uncertainty in the
  NIRC2 pixel scale in determining the error on the total mass.  The
  net effect is negligible, given the much larger ACS errors compared
  to the Keck LGS data and the fact that the error in the total mass
  is dominated by the parallax error (which is $15\times$ larger than
  the NIRC2 pixel scale uncertainty).}

As a consistency check, we also ran our MCMC fitting code on the
astrometric data for the binary L~dwarf 2MASSW~J0746+2000AB from
\citet{2004A&A...423..341B}.  We found excellent agreement between the
orbital parameters derived by us (using MCMC) and by Bouy \etal\
(using a variety of chi-square minimization approaches).  Not only do
the results agree to within the quoted errors, there is better than
$\approx$1\% agreement on the best-fit results and better than
$\approx$20\% agreement on the 95\% confidence intervals.


\subsubsection{Fitting Results \label{sec:fitting}}

\begin{figure}
\vskip -1in
\centerline{\includegraphics[height=7.5in,angle=90]{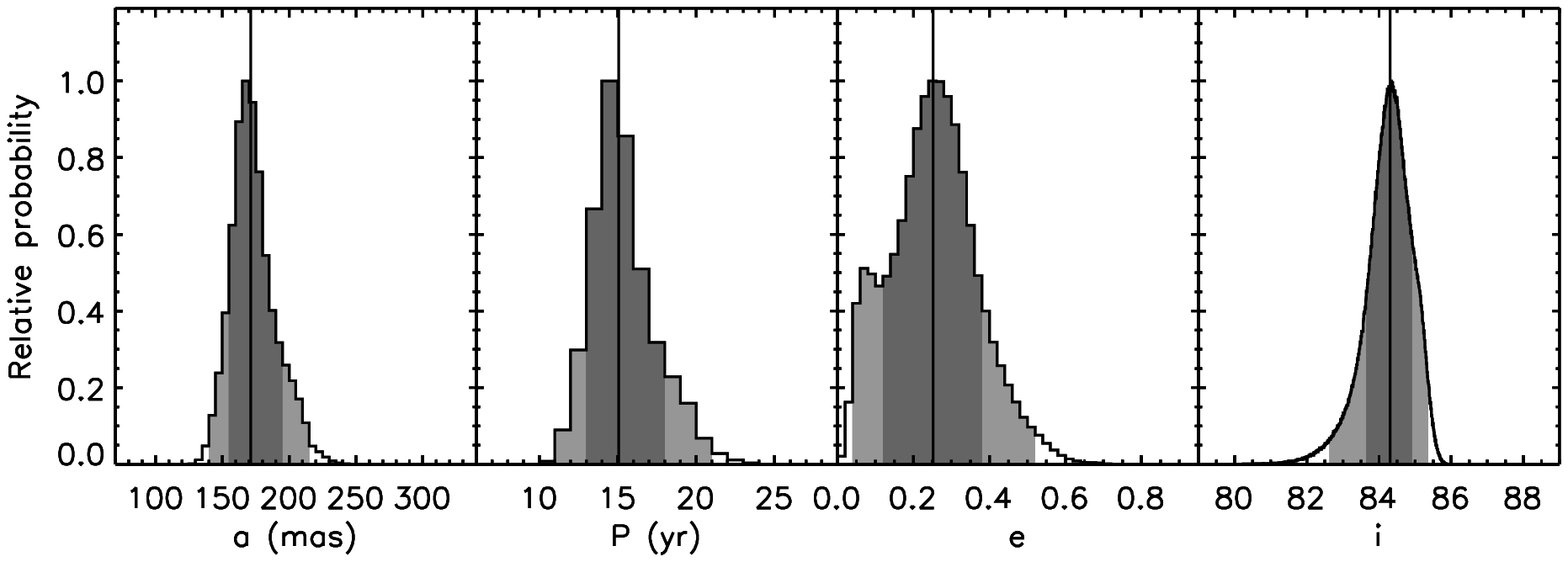}}
\centerline{\includegraphics[height=7.5in,angle=90]{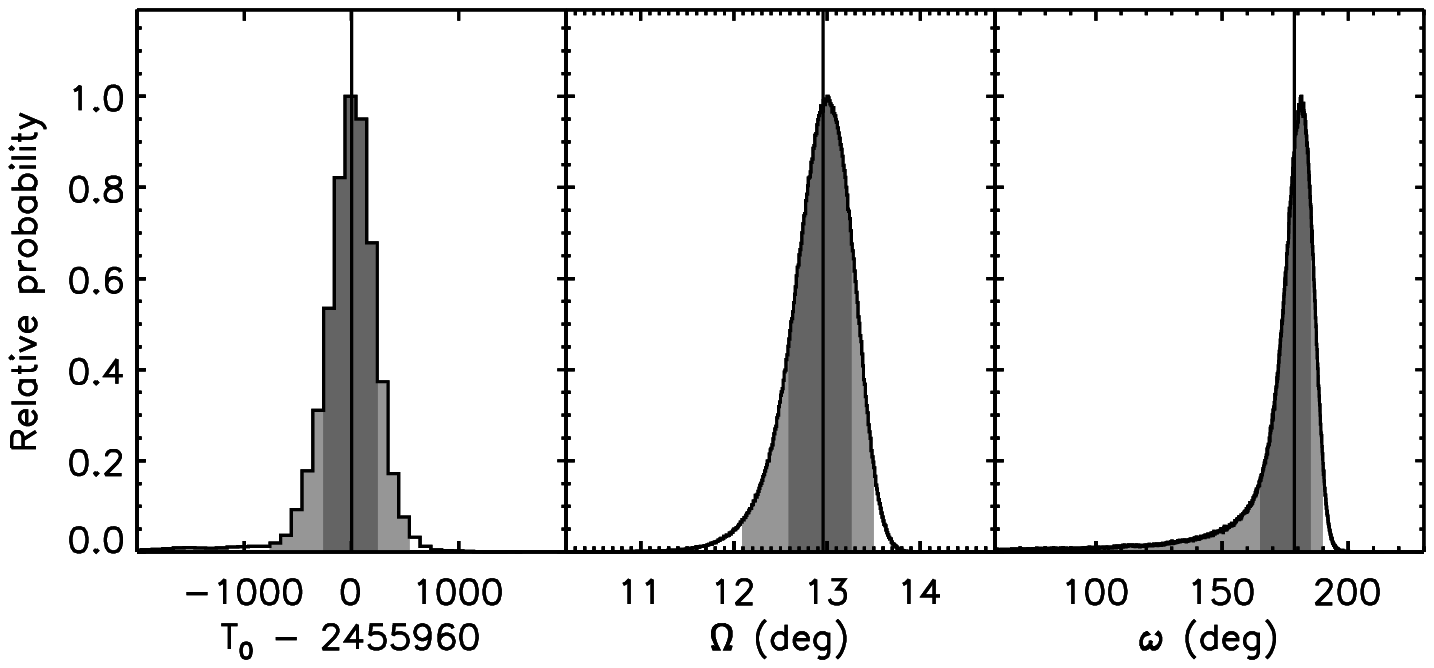}}
\caption{\normalsize Resulting MCMC probability distributions of
  orbital parameters: semi-major axis ($a$), orbital period ($P$),
  eccentricity ($e$), inclination ($i$), epoch of periastron ($T_0$),
  PA of the ascending node ($\Omega$), and argument of periastron
  ($\omega$).  Each histogram is shaded to indicate the 68.3\% and
  95.5\% confidence regions, which correspond to 1 and 2$\sigma$ for a
  normal distribution, and the solid vertical line represents the
  median value.\label{fig:orbit-summary}}
\end{figure}

Figures~\ref{fig:orbit-summary} show the resulting probability
distributions for the orbital parameters from the MCMC chain.  The
probability distributions are clearly not gaussian.
For a given parameter, we adopt the median as the result and describe
a confidence limit of $X$\% as simply the \hbox{$\case{1}{2} \pm
  \case{X}{200}$} bounds
of the sorted sample.  
At 68(95)\% confidence, we find a modest eccentricity of
0.25$_{-0.13(0.20)}^{+0.11(0.25)}$, an orbital period of
15.1$_{-1.6(3.1)}^{+2.3(5.1)}$~yr and a semi-major axis of
171$_{-13(27)}^{+19(41)}$~mas (2.3$_{-0.2(0.4)}^{+0.3(0.6)}$~AU
including the uncertainties in the plate scale and parallax).
Two of the orbital angles are very well-constrained, the inclination
$i = 84.3_{-0.6(1.7)}^{+0.6(1.0)}$~deg (nearly edge-on) and the PA of
the ascending node $\Omega = 13.0_{-0.3(0.9)}^{+0.3(0.5)}$~deg.  The
final results are summarized in Table~\ref{table:orbits}.

\begin{figure}
\vskip -1in
\hskip -0.3in
\centerline{\includegraphics[height=7.5in,angle=90]{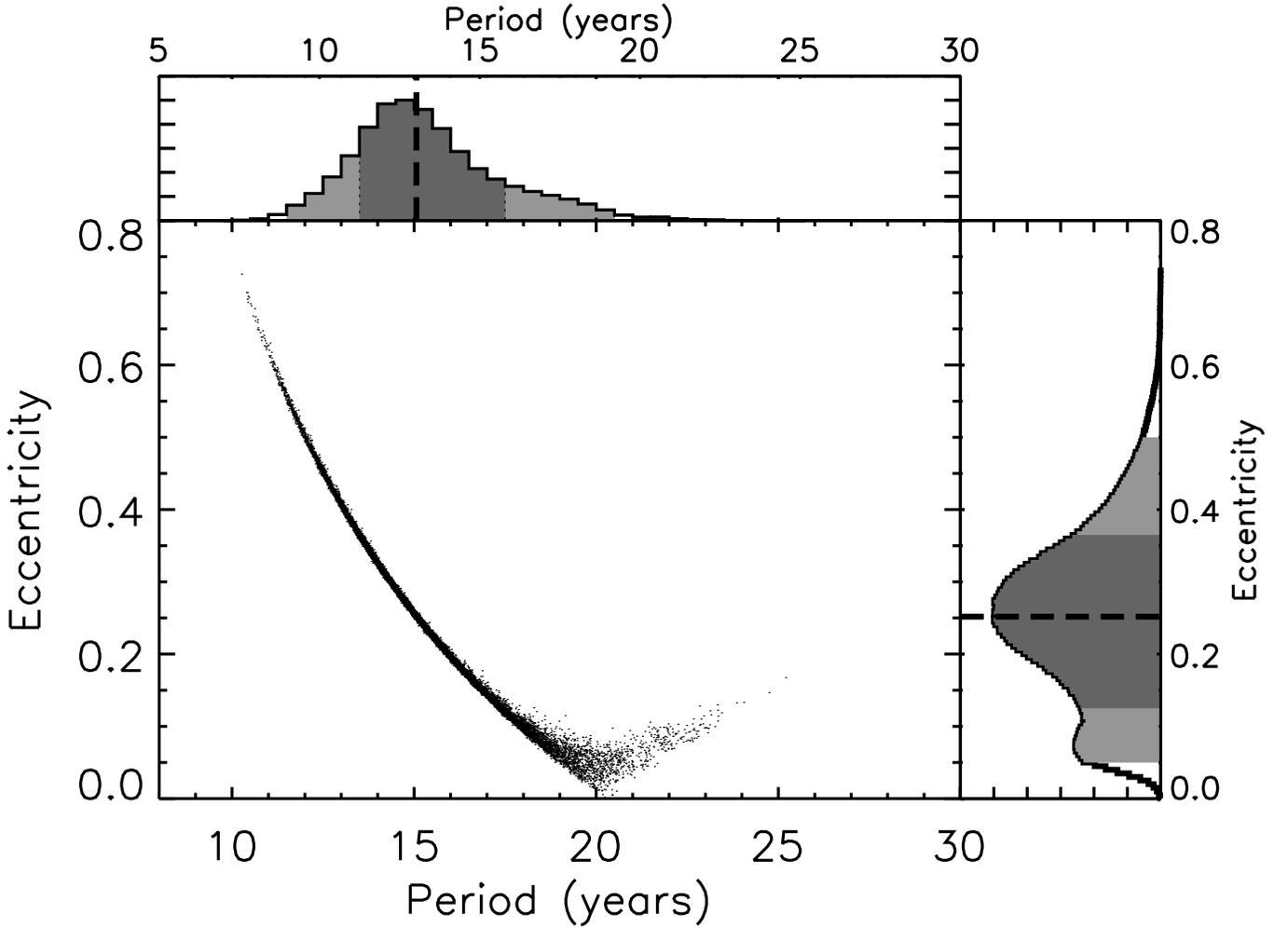}}
\vskip -2ex
\caption{\normalsize Results from MCMC determination of the orbital
  period and eccentricity for \twomassbin, illustrating the
  degeneracy between the two parameters.  The central plot shows
  all the values in the MCMC chain.  Two branches of possible orbits
  are seen, a short-period ($P<20$~yr) branch and a long-period 
  ($P>20$~yr) one.  About 98\% of the MCMC chain steps are in the short-period branch. The top
  and side plots show the resulting probability distributions of $P$
  and $e$.  Each histogram is shaded to indicate the 68.3\% and 95.5\%
  confidence limits, which correspond to 1$\sigma$ and 2$\sigma$ for a
  normal distribution, and the dashed vertical lines represent the
  median values.  
  \label{fig:orbit-p_e}}
\end{figure}

\begin{figure}
\vskip -1in
\hskip -0.3in
\centerline{\includegraphics[height=7.5in,angle=90]{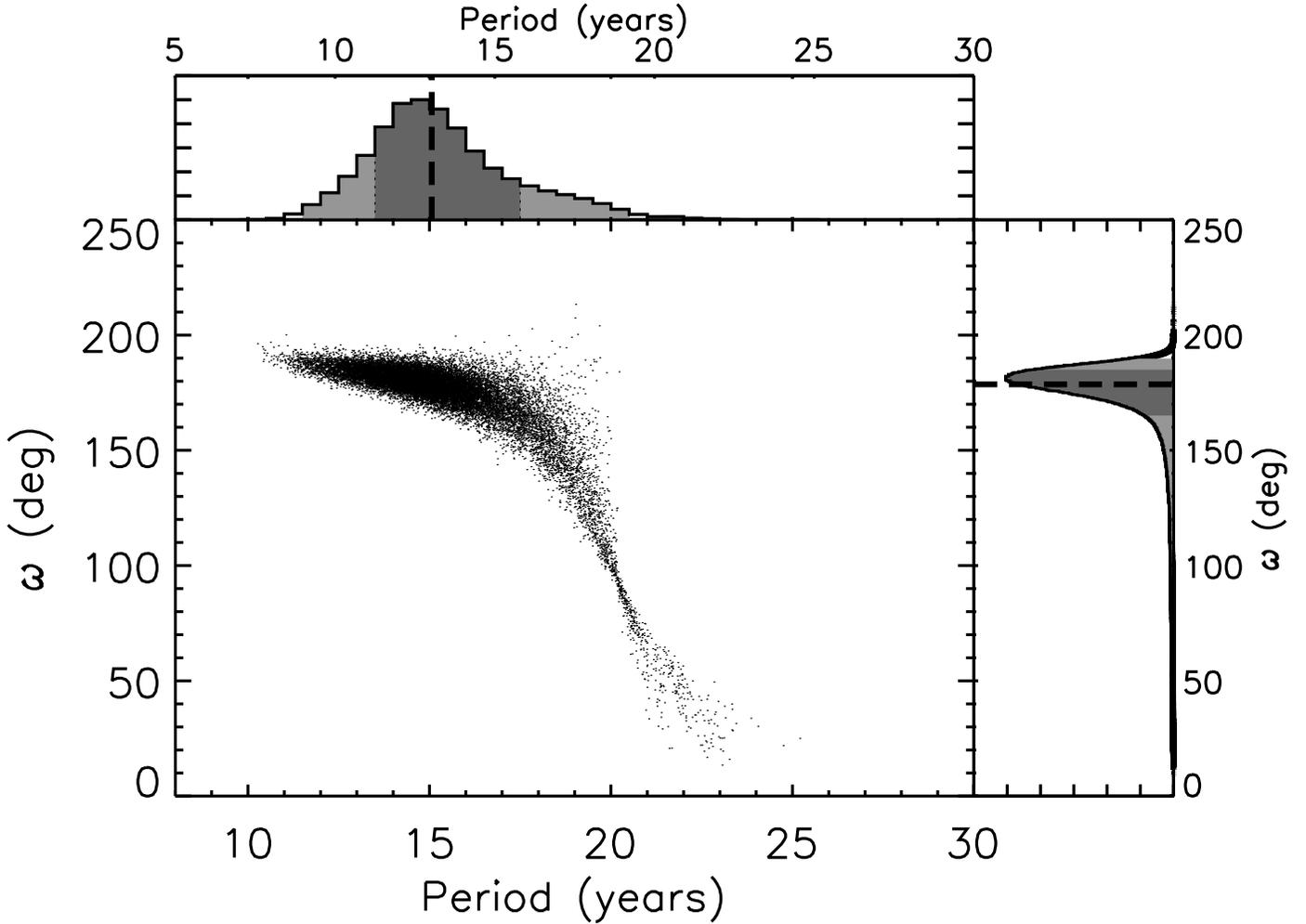}}
\vskip -2ex
\caption{\normalsize Results from MCMC determination of orbital period
  and the argument of periastron illustrating the degeneracy between
  these two parameters.  The locus illustrates the two general
  branches of possible orbits, with short-period orbits having just
  completed apoastron and long-period orbits having just completed
  periastron.  See Figure~\ref{fig:orbit-p_e} for further
  explanation. \label{fig:orbit-p_omg}}
\end{figure}

Figure~\ref{fig:orbit-p_e} shows the strong correlation between the
determination of the orbital period and the eccentricity.  It
illustrates that there are two classes of possible orbits: one branch
having shorter periods and smaller semi-major axes and the other
branch having longer periods and larger semi-major axes.  
Figure~\ref{fig:orbit-p_omg} shows that the short-$P$ branch orbits
have just passed apoastron ($\omega\approx 179\degs$, $\Omega \approx
13\degs$).  This is the favored solution, with 98\% of the steps in
the MCMC chain residing in this branch (using $e=0$ as the dividing
criteria in the $P-e$ plane).  However, a nearly circular orbit means
it can be difficult to clearly distinguish whether the system has just
passed apoastron or periastron, and thus a minority of the MCMC steps
(2\%) fall into the long-period branch.\footnote{Our inferred orbital
  period distribution is significantly longer than the 4-year estimate
  of \citet{2003ApJ...586..512B}.  Their original estimate was based
  on the projected physical separation at the discovery epoch, an
  assumed total mass of 0.07~\Msun, and the statistical estimate from
  \citet{1992ApJ...396..178F} that the true semi-major axis is on
  average 1.26$\times$ larger than the projected separation.  The
  large discrepancy with our orbital period determination is not
  surprising, since a single epoch of imaging provides a highly
  uncertain period estimate --- the likely true periods can span a
  factor of several greater or smaller
  \citep[e.g.][]{1999PASP..111..169T}.}


The MCMC fitting provides probability distributions for the orbital
parameters, but does not provide a single best-fitting orbit per~se,
since a range of possible orbits fit the data with similar \rchisq\
values.  One illustration of this is the result for $\omega$ =
179$_{-14(83)}^{+6(11)}$~deg, where the 95\% confidence limits are
broad enough to span both the short-period and long-period solution
branches.  Thus to plot orbits on the sky, we employ gradient-descent
methods to find the best-fitting orbit with the MCMC-derived values as
the starting point.  Figures~\ref{fig:orbit-orbit}
and~\ref{fig:orbit-sep} shows the resulting orbit, which has a period
of 15.2~yr, a total mass of 0.0556~\Msun\ and $\rchisq=0.9$.  To
illustrate how the uncertainty in the orbital period impacts the
orbit, we also show the best-fitting orbits found when fixing the
period to 12 and 20~yr, which have total masses of 0.0523 and
0.0590~\Msun\ and $\rchisq$ of 1.1 and 1.0, respectively.  All three
orbits reside in the short-period branch and show that the projected
separation is now rapidly decreasing.  The system is expected to be
well-resolved again in the year 2011.


Applying Kepler's Third Law to the period and semi-major axis
distributions gives the posterior probability distribution for the
total mass of the binary, with median of 0.0556~\Msun, a standard
deviation of 0.0018~\Msun\ (3.2\%), and a 68(95)\% confidence range of
about $\pm$0.0018(0.0037)~\Msun\
(Figure~\ref{fig:orbit-masses}).
However, the MCMC probability distribution does not include the
uncertainties in the parallax (1.6\%) and the NIRC2 pixel scale
(0.11\%).  By Kepler's Third Law, the quadrature sum of these errors
amounts to an additional 4.9\% uncertainty on the derived total mass.
Since the MCMC-derived mass distribution is asymmetric, we account for
this additional error in a Monte Carlo fashion; for each step in the
chain, we draw a value for the pixel scale and parallax from a normal
distribution and then compute the total mass.  The resulting mass
distribution is essentially gaussian (Figure~\ref{fig:orbit-masses}).
Our final determination of the total mass is $0.056 \pm 0.003
(0.006)$~\Msun\ at 68(95)\% confidence.  Thus, the total mass of this
system is well-measured, with the parallax error being the dominant
uncertainty.  This is the coolest and lowest mass binary with a
dynamical mass determination to date.


\subsubsection{Alternative Orbit Fits \label{sec:altorbits}}

The WPFC2 discovery epoch in 2000 is obviously a key component to
fitting the orbit.  As described in \S~\ref{sec:wfpc2}, we
independently analyzed this dataset and greatly reduced the
measurement errors compared those reported by
\citet{2003ApJ...586..512B}.  To examine the impact of this
improvement, we also tried fitting the orbit using the original
Burgasser \etal\ astrometry.  Without the 2008 Keck data, our improved
WFPC2 astrometry is essential for a well-constrained fit.  However,
with the complete dataset, the fitted orbital parameters and the total
mass are insensitive to the specific choice of WFPC2 astrometry,
changing by less than 1$\sigma$.

Also, to check the effect of our assumed prior on the MCMC fitting
(flat in $\log P$ and $\log a$), we tried three alternative priors
from the literature:

\begin{enumerate}

\item {\em Solar-type stars:} \citet{1991A&A...248..485D} analyzed a
  well-defined sample of 164 nearby solar-type stars (spectal type F7
  to G9) and found a log-normal distribution in orbital period:
\begin{equation}
\frac{dN}{d(\log P)}\sim \exp\left[\frac{-(\log P -
      4.8)^2}{2\sigma_{\log P}^2}\right]
\end{equation}
where $P$ is the period in days and $\sigma_{\log P} = 2.3$.

\item {\em Ultracool visual binaries:} \citet{2007ApJ...668..492A}
  conducted a detailed analysis of published imaging surveys of 361
  ultracool field objects to model the separation distribution as a
  log-normal distribution:
\begin{equation}
\frac{dN}{d(\log a)} \sim \exp\left[\frac{-(\log a -
      0.86)^2}{2\sigma_{\log a}^2}\right]
\end{equation}
where $a$ is the semi-major axis in AU and $\sigma_{\log a} =
0.28$.\footnote{Since this distribution is derived from analysis of
  ultracool field dwarfs, one might arguably choose this as the
  default prior.  However, Allen's input data are restricted to
  imaging and do not include any spectroscopic binaries; therefore,
  the smallest semi-major axes are poorly constrained.  For instance,
  the Allen distribution predicts basically no binaries at 1~AU or
  smaller separations ($\gtrsim3\sigma$ events), perhaps at variance
  with spectroscopic binary studies \citep{2005MNRAS.362L..45M,
    2006AJ....132..663B}.  The young eclipsing M6.5+M6.5 binary
  2MASS~J0535$-$05 \citep{2006Natur.440..311S} is also highly
  anomalous with this distribution ($\sim8\sigma$ event).  Finally,
  the Allen analysis does not separately consider objects of different
  spectral types, whereas \citet{2006ApJS..166..585B} have suggested
  the separation distribution for T~dwarfs may be tighter than for the
  L~dwarfs.  Thus our default prior of flat in $\log a$ is a
  conservative choice.  Note that longer period of \twomassbin\ we
  find relative to the original estimate of $\sim$4~years
  ($a\sim1.1$~AU) is in accord with the distribution proposed by
  Allen, which would indicate that such a $\sim$1.1~AU system would be
  uncommonly rare ($\approx3\sigma$) relative to the $\sim$100 known
  ultracool binaries.}

\item {\em Ultracool visual and spectroscopic binaries:}
  \citet{2005MNRAS.362L..45M} analyzed a sample of 47 ultracool
  binaries with multiple radial velocity measurements and adopted a
  log-normal distribution truncated at large separations to match the
  known visual ultracool binaries:
\begin{equation}
  \frac{dN}{d(\log a)} \sim \exp\left[\frac{-(\log a -
      0.6)^2}{2\sigma_{\log a}^2}\right] \qquad \hbox{\rm for $a < 10$~AU}
\end{equation}
where $a$ is the semi-major axis in AU, $\sigma_{\log a} = 1.0$, and
$dN/d(\log a) = 0$ for $a > 10$~AU.

\end{enumerate}

Figure~\ref{fig:priors} shows the posterior probability distributions
for the total mass from the different priors.  Overall, the choice of
prior has very little effect on the mass determination.  The
\citet{2007ApJ...668..492A} distribution favors slightly higher
masses, but its results are consistent with the other priors.

As a final independent check, we also fit our astrometry using the
linearized least-squares fitting routine ORBIT
\citep{1999A&A...351..619F}, using the MCMC-derived parameters as the
starting guess.  The ORBIT results are given in
Table~\ref{table:orbits}, with a resulting $\rchisq=0.9$, and show
excellent agreement with the MCMC results.


\section{Discussion}

A primary goal of measuring fundamental properties for ultracool
binaries is to compare the measurements against theoretical models of
their physical properties.  A number of studies have been published
for the previous ultracool visual binaries with dynamical masses
(\S~1), with subtle and/or overt differences in the ways that
observations are compared to models.  In the analysis that follows, we
strive to clearly elucidate the comparison of our \twomassbin\
observations to the models, both in terms of its approach and
limitations.  From the standpoint of the observations, we have high
quality measurements of (1) the total mass of the system and (2) the
individual absolute magnitudes, with (3) the individual bolometric
luminosities only slightly less reliable.  (We have not measured the
complete spectral energy distribution but have accounted for this
uncertainty in computing \Lbol\ in \S~\ref{sec:lbol}).  We now examine
what can be learned from these data in concert with evolutionary
models and theoretical atmospheres.

\subsection{Substellarity}

The most immediate result from our measurement is that \twomassbin\ is
a bona fide brown dwarf binary.  The total mass of
$0.056\pm0.003$~\Msun\ is below the solar-metallicity
stellar/substellar boundary of $\approx$0.070--0.074~\Msun\
\citep[e.g.][]{1963PThPh..30..460H, 1963ApJ...137.1121K, bur01}, with
the boundary increasing to higher masses for lower metallicities
\citep{1994ApJ...424..333S} --- therefore, the individual components
are clearly substellar.  This is the second binary where both
components are directly confirmed to be brown dwarfs, after the young
eclipsing M6.5+M6.5 binary 2MASS~J0535$-$05
\citep{2006Natur.440..311S}, and is the first such field binary in
this category.

\subsection{Age \label{sec:ages}}

Brown dwarfs follow a mass-luminosity-age relation.  We have measured
two of these quantities, the (total) mass and the luminosity, and by
using evolutionary tracks we can determine the third quantity, the age
of the system.  We use models from the Tucson group
\citep{1997ApJ...491..856B}, which provide predictions for \Lbol, and
the ``COND'' models from the Lyon group \citep{2003A&A...402..701B},
which predict both \Lbol\ and absolute magnitudes.  We conservatively
assume the system is coeval and that the system is a true binary, not a
partially (un)resolved higher order multiple system (\S~\ref{sec:phot}).

For each tabulated model age, we use the individual absolute
magnitudes and/or bolometric luminosities to calculate the mass of the
components and then sum the masses.  We then apply the observed total
mass to determine the age range of the system.  All measurement errors
in \Lbol\ and the total mass are accounted for in a Monte Carlo
fashion, namely we repeat the model calculations over multiple
realizations for the \Lbol\ and the total mass values.  We take great
care to account for the covariance between the relevant quantities in
the calculation.  For instance, the total mass of the system and the
luminosity both depend on the parallax, and thus their errors are
positively correlated; we therefore draw the parallax values from a
normal distribution and incorporate these in determining the Monte
Carlo distribution of total masses and luminosities, which themselveed
are then propagated in the model-based calculations.  This approach
results in a probability distribution for the system's age (as well as
the other resulting parameters discussed below), which we summarize
with the median value and confidence limits.

Figure~\ref{fig:models-totalmass} shows the results of these
calculations to determine the age of the system.  For a consistent
comparison between the Lyon and Tucson models, we use only the results
derived from the \Lbol\ measurements. However, the Figure also shows
that using the absolute magnitudes predicted by the Lyon models would
give similar results.

We determine an age of $0.73 \pm 0.07(0.15)$ Gyr from the Burrows
models and $0.83 \pm 0.08(0.18)$ Gyr from the Baraffe models at
68(95)\% confidence.  To construct a representatve ``average'' of the
model results, we merge the results of the individual Monte Carlo
calculations into a single distribution and compute its confidence
limits.  Thus, we assign an age of \hbox{$0.78\pm0.09(0.18)$~Gyr}
(Table~\ref{table:evolmodels}).  This is relatively youthful compared
to the main-sequence stars in the solar neighborhood, \eg,
$\approx$95\% of nearby solar-type stars have age estimates of
$\gtrsim$1~Gyr \citep{2004A&A...418..989N}.  However, the mean age of
T~dwarfs is expected to be younger than for field stars, since the
known census is magnitude-limited and younger objects are brighter.
The age distribution of field ultracool dwarfs has been modeled by
\citet{2004ApJS..155..191B} and \citet{2005ApJ...625..385A}; they
generally find that field T~dwarfs can span younger ages than for
low-mass stars, though the predicted age distributions for both types
of objects have large spreads.

Kinematics provide an independent (albeit indirect) indicator of age,
as older objects are expected to generally show larger space motions
due to their accumulated history of dynamical interactions
\citep[e.g.][]{1977A&A....60..263W}.  The tangential velocity of
\twomassbin\ ($V_{tan} = 17.3 \pm 0.4~\kms$) is the second smallest
measured for T~dwarfs, with only the T5.5~dwarf 2MASS~1546$-$3325
being smaller ($12.1\pm0.4~\kms$; \citealp{2003AJ....126..975T}).
This is generally in accord with the $0.78\pm0.09$~Gyr inferred from
the evolutionary models, namely that \twomassbin\ is among the
youngest members of the nearby field population.  However, since the
measured $V_{tan}$ distribution of field T~dwarfs is quite broad, with
an unweighted average of 38.4~\kms\ and a standard deviation of
20.4~\kms\ among the 21 unique objects in the
\citet{2003AJ....126..975T} and \citet{2004AJ....127.2948V} parallax
samples, \twomassbin\ does not appear to be anomalously young for a
field object.  A radial velocity measurement is needed to determine
the binary's space motion and thus better constrain its kinematics.

\subsection{Temperatures and Surface Gravities  \label{sec:temp}}

With the age of the system determined above, the combination of the
observations and the evolutionary models provide highly precise values
for the remaining physical parameters.  The results derived from the
two sets of evolutionary tracks are given in
Table~\ref{table:evolmodels} and are computed from the same Monte
Carlo approach that accounts for the covariance in the measurements.
The Tucson and Lyon models give consistent values, with the Tucson
models giving slightly larger radii and thus slightly cooler
temperatures.  Again, to compute a representative ``average'' for each
parameter, we merge the Monte Carlo distributions computed from each
set of models and compute confidence limits for the aggregate.  This
is not intended to be physically meaningful, but rather is a
quantitative representation of the results that accounts for
non-gaussian and/or inconsistent distributions from the two sets of
models.
We thus find radii of \hbox{$0.0986 \pm 0.0015~\Rsun$} and
\hbox{$0.0993 \pm 0.0017~\Rsun$}, effective temperatures of
$1028\pm17$~K and $978\pm17$~K, and surface gravities of \hbox{$\logg
  = 4.91 \pm 0.04$} and $4.87 \pm 0.04$ for components~A and~B,
respectively.\footnote{Since all these derived properties rely on the
  model-derived age, their distributions from the Monte Carlo
  calculations are fairly correlated.  However, the formal confidence
  limits on the quantities are so small that this correlation is
  unlikely to be significant for any future analyses.}

To re-iterate, these properties are derived using only the measured
{\em total} mass and resolved magnitudes/luminosities, along with the
assumption that the system is coeval and composed of two components.
No additional assumptions have been made to determine the individual
masses.  We also have avoided using spectral types and/or effective
temperatures in this aspect of our analysis (contrary to some previous
studies), as these quantities can introduce additional systematic
errors and/or circular reasoning.  For instance, it would be incorrect
to employ the relations between spectral type and \Teff\ from
\citet{gol04} or \citet{2004AJ....127.2948V} to determine \Teff\ for
the two components and then compare to evolutionary tracks, as the
Vrba \etal\ and Golimowski \etal\ relations are derived from the radii
of field brown dwarfs predicted by the evolutionary tracks themselves.
Likewise, it is not necessary to use \Teff\ determinations from
atmospheric models to determine the age of the system or the physical
properties of the individual components in our approach.  (See also
\S~\ref{sec:tests}.)

As already noted above, our Monte Carlo calculations account for the
covariance in the measurements.  One important effect is that the
uncertainties derived from the resolved magnitudes and luminosities of
the two components are correlated, since they all depend on the
measurement uncertainties in the integrated-light photometry of the
system.  As a consequence, the {\em relative} temperture difference
between the two components ($\Delta\Teff = 50_{-10}^{+6}$~K) can be
calculated to higher precision than would be indicated by the
uncertainties in the individual \Teff\ determinations ($\sqrt{2}
\times 17$~K = 24~K).  This $\Delta\Teff$ agrees with the
$\approx$70~K difference expected from the \citet{gol04} polynomial
fits for the 0.5~subclass difference between the two components.  (The
$\Delta\Teff$ from the Golimowski \etal\ fits would be about twice as
large for a 1~subclass difference.)

The physical parameters for 2MASS~J1534$-$2952A and~B are in general
agreement with previous determinations for the properties of field
T~dwarfs.  However, our values have much higher precision, because the
accurate total mass measurement leads to a small age range, which
leads to strong constraints on the radii and thus small uncertainties
on the derived \Teff\ and \logg\ values.  (We discuss this further in
\S~\ref{sec:tests}.)  T~dwarf surface gravities have been inferred to
be $\log(g) = 4.5-5.5$ by comparing theoretical model atmospheres to
optical spectra \citep{2002ApJ...573..394B}, near-IR colors and line
strengths \citep{2004AJ....127.3553K}, and low-resolution
near-IR/mid-IR spectra \citep{2006ApJ...639.1095B,
  2006astro.ph.11062S, 2007arXiv0705.2602L, 2007arXiv0711.0801C}.
This range encompasses our \logg\ determinations for \twomassbin.

On the other hand, our very precise temperatures for the two
components of \twomassbin\ are discrepant with previous studies of
T~dwarfs: the \Teff's of \twomassbin\ appear to be cooler than
determined previously for mid-T~dwarfs.  These discrepancies occur for
two {\em separate} comparisons.


\subsubsection{Temperature discrepancy with evolutionary models}

Temperatures for field T~dwarfs have been inferred by combining
accurate \Lbol\ determinations with radius predictions from
evolutionary tracks.  This approach is expected to be reasonably
accurate, since the radii of brown dwarfs older than $\sim$100~Myr are
predicted to vary by $\lesssim$30\%.  \citet{gol04} adopt a typical
age of 3~Gyr and a plausible range of 0.1--10~Gyr in computing \Teff\
from \Lbol, and \citet{2004AJ....127.2948V} adopt a radius range of
$0.90\pm0.15~\Rjup$ predicted by \citet{2002PhDT........27B}
simulations of the solar neighborhood assuming a constant star
formation history.  (Both studies use the Tucson evolutionary tracks.)
There are 8~T4.5--T6.5 dwarfs in these studies (after updating the
spectral types to the latest classification by
\citet{2005astro.ph.10090B}), and all of these appear to be single
based on high angular resolution imaging
(\citealp{2003ApJ...586..512B, 2006ApJS..166..585B}; Liu \etal, in
prep).  We compute the average and standard deviation of both studies
to obtain $\Teff = 1216\pm20$~K for T4.5 (2~objects), 1146~K for T5.0
(interpolated), 1077~K for T5.5 (1~object), $1014\pm33$~K for T6.0
(3~objects), and $950\pm106$~K for T6.5 (3 objects).  We have excluded
the T6+T8 binary 2MASS~J1225$-$2739AB and the peculiar T6 dwarf
2MASS~J0937+2931, and we have assumed the T4.5~dwarf
2MASS~J0559$-$1404 is an equal-magnitude unresolved binary based on
its pronounced overluminosity (\eg, Figure~3 of
\citealp{burg2006-lt}).  Neither sample contains any T5.0 objects, but
the 4~other subclasses almost exactly follow a straight line, so we
linearly interpolate to find \Teff\ for T5.0.

  In comparison to the field objects, the components of \twomassbin\
  appear to have $\approx$100~K cooler temperatures relative to their
  spectral subclass ($\approx120\pm35$~K for the primary and
  $\approx100\pm35$~K for the secondary, where we have adopted a 30~K
  uncertainty for the T5 and T5.5 field objects based on the other
  subclasses with more than one object).  The disagreement is modest,
  and a more definitive comparison is hampered by the few \Teff\
  determinations (\ie, parallaxes) for T4.5--T6.5 dwarfs.
  Nevertheless, the result is potentially intriguing.

  In particular, \citet{2006astro.ph..7514M} have noted perhaps a
  similar effect for the three known late-L (L7--L8) dwarf companions
  to field stars.  More precise \Teff\ estimates can be obtained from
  evolutionary models for these companions than for field objects by
  incorporating the age estimates of their primary stars.  (See
  \S~\ref{sec:tests} for details.)  Metchev \& Hillenbrand find that
  the L~dwarf companions appear to be $\approx$100--200~K cooler than
  single field late-L dwarfs.  They raise the possibility that the
  model-derived radii are at fault, either due to incorrect cooling
  rates or systematic overestimate of the field dwarf ages.  However,
  they prefer the hypothesis that the discrepancy is a manifestation
  of an unanticipated surface gravity dependence of the L/T
  transition, causing younger L/T transition objects to have cooler
  temperatures than older ones.  This is motivated by their analysis
  of the young (0.1--0.4~Gyr) L7.5 companion HD~203030B and apparently
  supported by the young (0.1--0.5~Gyr) T2.5 companion HN~Peg~B, which
  also appears to be $\approx$200~K cooler than field objects of the
  same spectral type \citep{2006astro.ph..9464L, 2008arXiv0804.1386L}.

  We find that the T5.0 and T5.5 components of \twomassbin\ may also
  be $\approx$100~K cooler than comparable field objects.  Since these
  two components are later-type than the L/T transition (\eg, their
  positions in IR color-magnitude diagrams is coincident with the
  locus of mid/late-T dwarfs with blue near-IR colors), this may
  suggest that the \Teff\ discrepancy might not be solely associated
  with the L/T transition.  Instead, \twomassbin\ and the
  aforementioned L/T companions may indicate that a systematic error
  in the estimated ages and radii of field late-L and T~dwarfs is the
  culprit.

  The $\approx$10\% temperature discrepancy for \twomassbin\ amounts a
  $\approx$20\% underestimate of the radii.  For a fixed value of
  \Lbol\ (which is the appropriate constraint here), the \citet{bur01}
  scaling relations give
  \begin{equation}
    t \sim R^{-8.56}
  \end{equation} 
  where $t$ is the age and $R$ is the radius.  This agrees well with
  the exponent value of 8.2--8.4 extracted from the Tucson models for
  sources of $\log(\Lbol/\Lsun) = -5.0$ to $-5.1$.  Propagating the
  $\pm3\%$ uncertainties in the \Teff\ disagreement, the implied age
  overestimate is a factor of $6\pm3$, meaning implied ages of
  $\approx$0.3--1.0~Gyr for the field population.\footnote{Applying
    the same scaling relation to the 10--15\% radius discrepancy found
    by \cite{2006astro.ph..7514M} for objects at the L/T transition
    implies a factor of 2--3 overestimate in the representative age of
    the field population.}  

  The same discrepancy can be seen in an alternate fashion, namely by
  comparing the luminosities for the same T4.5--T6.5 field objects:
  $\log(\Lbol/\Lsun) = -4.79\pm0.04$ for T4.5, $-4.89$ for T5.0
  (interpolated), $-4.99\pm0.01$ for T5.5, $-5.17\pm0.29$ for T6.0,
  and $-5.22\pm0.20$ for T6.5.  The luminosities of \twomassbin\ are
  comparable to the field objects of similar type ($-5.015\pm0.019$
  for T5.0~component~A and $-5.093\pm0.019$ for T5.5~component~B).
  Thus, in order for all the objects to have similar temperatures and
  \Lbol, they must have about the same radius and thus about the same
  age as \twomassbin.  In other words, the measured total mass of
  \twomassbin\ is too small (by a factor of $\approx$2) compared to
  the mass expected from the evolutionary models for 3~Gyr objects
  with $\log(\Lbol/\Lsun)\approx-5.0$.

  A representative age of $\sim$0.5~Gyr for the field population is
  not ruled out given the state of the observations.  Though
  Golimowksi \etal\ did consider the range of 0.1--10~Gyr, they
  adopted a nominal age of 3~Gyr in determining \Teff\ for field
  dwarfs, based on the 2--4~Gyr age estimate from the tangential
  velocities of ultracool dwarfs by \citet{2002AJ....124.1170D}.  A
  younger age could be accomodated, since the tangential velocity of a
  population is only an approximate statistical estimate of its age.
  Indeed, kinematic analysis of the space motions of L and T~dwarfs
  suggests a younger age of $\approx$0.5--2~Gyr
  \citep{2007ApJ...666.1205Z}.  Similarly, the radii of
  $0.90\pm0.15~\Rjup$ adopted by Vrba \etal\ is based on a mass
  function where $dN/dM \sim M^{-1}$; a somewhat steeper mass function
  would lead to younger typical ages (\eg, Figure~8 of
  \citealp{2004ApJS..155..191B}, though \citealp{2007arXiv0710.4157M}
  suggest $dN/dM \sim M^{0}$ based on a small sample of T~dwarfs).
  Thus, the discrepancy of evolutionary model-derived temperatures
  between objects of known mass/age and the field population can be
  plausibly explained by a modest overestimate of the ages of the
  field population.  A larger sample of ultracool dwarfs with known
  masses and/or ages is needed to better explore this issue
  (\S~\ref{sec:tests}).

  \subsubsection{Temperature discrepancy with model atmospheres}

  The spectrum of \twomassbin\ has not yet been fitted with model
  atmospheres due to its composite nature.  In fact, the T~dwarf class
  as a whole has not been extensively subjected to such comparisons.
  \citet{2006ApJ...639.1095B} determined \Teff\ for a sample of
  sixteen T5.5--T8 dwarfs by comparing near-IR spectral indices to
  condensate(dust)-free atmosphere models from the Tucson group.  They
  determined $\Teff = 1020-1100$~K for one T5.5 dwarf, and an
  unweighted linear fit of atmosphere-derived \Teff\ versus spectral
  type for their sample (excluding the peculiar T6~dwarf
  2MASS~J0937+2931) gives
  \begin{equation}
    \Teff_{\rm,\ atmosphere} = 1090 - 126 \times (SpT-25.5)
  \label{eqn:teff-atm}
  \end{equation}
  where $SpT=25.5$ for T5.5, $SpT=26$ for T6, etc.  The RMS about the
  linear fit is 50~K, which we adopt as the uncertainty
  (a value somewhat larger than the $\pm$10~K to $\pm$40~K range
  computed for individual objects in their sample).
  Extrapolating the linear fit gives $\Teff = 1160$~K for T5.  This is
  obviously approximate, \eg, given the potential systematic effects in
  the models and the spectral classification scheme, though this value
  agrees with the $\Teff=1150-1200$~K found by by fitting model
  atmospheres to the \hbox{0.95--14.5~\micron} spectrum of the
  T4.5~dwarf 2MASS~J0559$-$1404 \citep{2007arXiv0711.0801C}.

  Thus, model atmospheres indicate $\Teff = 1160\pm50$~K and
  $1090\pm50$~K for 2MASS~J1534$-$2952A and~B, respectively.  The
  temperatures we find using evolutionary tracks appear to be cooler by
  $\approx$100~K at modest significance ($-130\pm50$~K for component~A
  and $-110\pm50$~K for component~B, if we assume that the errors add in
  quadrature).  We cannot objectively discern if the problem lies in the
  evolutionary tracks, the model atmospheres, or both.  However, the
  evolutionary models are thought to be robust to the principal input
  uncertainties \citep{2000ApJ...542..464C}.  On the other hand, the
  model atmospheres are quite uncertain.  Even though the spectral
  appearance of mid- and late-T dwarfs is relatively simple ---
  dominated by collision-induced \htwo, \htwoo, and \meth\ in the
  near-IR and the wings of the \ion{K}{1} 0.77~\micron\ resonance line
  in the far-red --- the line lists for \htwoo\ and \meth\ are known to
  be incomplete, and the input physics to the atmosphere models are
  complex.  Current atmospheres generally match the observed spectra of
  late-T (T6--T8) dwarfs, but not exactly so
  \citep[e.g.][]{2005astro.ph..9066B, 2006ApJ...639.1095B,
    2006astro.ph.11062S, 2007arXiv0705.2602L}.\footnote{For instance,
    recognizing these limitations, \citet{2006ApJ...639.1095B} chose to
    calibrate the model atmosphere predictions empirically using the
    well-studied T7.5~dwarf Gl~570D in fitting models to late-T dwarf
    spectra, as opposed to using the atmospheres directly.}  Therefore,
  while a larger sample of objects is needed both for dynamical mass
  determinations and model atmosphere fitting, the plausible hypothesis
  is that the observed discrepancy arises from an overprediction of
  \Teff\ by current model atmospheres.


\subsection{Color-Magnitude and Hertzsprung-Russell Diagrams \label{sec:hrd}}

We have directly measured the total mass of the \twomassbin\ system.
However, using the evolutionary tracks to determine the physical
properties also implicitly determines the mass ratio, since the
model-derived age and observed individual luminosities translate into
individual masses (again with the assumption that the system is
composed of only two components).  We infer the mass ratio of the
system from the ratio of the bolometric luminosities, as this is very
robust.  To illustrate this, consider the analytic scaling relation
for solar-metallicity substellar objects from \citet{bur01}:
\begin{equation}
  \Lbol \sim M^{2.64}\ t^{-1.3}\ \kappa_R^{0.35}
\label{eqn:lbol}
\end{equation}
where $M$ is the mass, $t$ is the age and $\kappa_R$ is the Rosseland
mean opacity.  We measure a $0.078\pm 0.016$~dex difference in \Lbol\
between the two components\footnote{At face value, the results in
  Table~\ref{table:resolved} would give an uncertainty of $\sqrt{2}
  \times 0.018 = 0.025$~dex in the \Lbol\ difference, but this would
  include the uncertainties in the distance modulus (0.04~mag) and the
  integrated-light photometry (0.03~mag), which are common to both
  components.}
which leads to a mass ratio {$q \equiv M_B/M_A = 0.934\pm0.007$}.  The
uncertainty in the mass ratio is small due to the weak dependence of
mass on luminosity at fixed age.  Using the actual tabulated Tucson
and Lyon models and again keeping careful track of the covariance in
the calculations, we compute a final value of
\hbox{$q=0.936_{-0.008}^{+0.012}$}, where the error includes the
uncertainties in the model-inferred age and the observed \Lbol\
difference.  (The Tucson and Lyon models give basically identical
results for $q$.)  This gives individual masses of
$0.0287\pm0.0016$~\Msun\ ($30.1\pm1.7$~\Mjup) and
$0.0269\pm0.0016$~\Msun\ ($28.2\pm1.7$~\Mjup) for components~A and~B,
respectively.

We first compare the individual components against the COND
evolutionary models of the Lyon group, which provide predictions for
the absolute magnitudes and colors.  The model predictions are
generated for the CIT photometric system, so we transform our resolved
MKO photometry for \twomassbin\ to this system using the results of
\citet{2004PASP..116....9S}.  Figure~\ref{fig:cmd} shows that the
models are somewhat too red compared to the data.  This is not
surprising, as model atmospheres for T~dwarfs are known to be
deficient in the \meth\ and \htwoo\ opacities relevant at these
wavelengths \citep[e.g.][]{2007arXiv0705.2602L}.  The plotted COND
models are also computed only for solar-metallicity, and a non-solar
metallicity for \twomassbin\ would impact the colors and magnitudes
\citep[e.g.][]{2006liu-hd3651b, 2006astro.ph..9556B}.  Indeed, current
models do not exactly match the observed color-magnitude loci for
field T~dwarfs (\eg, Figure~8 of \citealp{2004AJ....127.3553K} and
Figure~7 of \citealp{2005astro.ph..9066B}).  Nevertheless,
\twomassbin\ will provide a strong test to for future models, since
the components' magnitudes, colors, and masses are very well-measured.

With the individual mass estimates and an independent determination of
\Teff, it is possible to directly test different evolutionary tracks
using the Hertzsprung-Russell (H-R) diagram.  We use the values of
$\Teff = 1160\pm50$~K and $1090\pm50$~K for 2MASS~J1534$-$2952A and~B,
derived in \S~\ref{sec:temp} from model atmosphere studies.
Figure~\ref{fig:hrd-masses} shows the individual components on the
H-R~diagram and compares these to the Tucson and Lyon evolutionary
tracks.  The locations of the two components disagree with both sets
of models (which agree very well between themselves).

The likely interpretation is that the temperatures from the model
atmospheres place the components to the left of the evolutionary
tracks, \ie, too warm.  As discussed in \S~\ref{sec:temp}, the model
atmospheres are a significant source of the uncertainty in placing the
components on Figure~\ref{fig:hrd-masses}.  A possible systematic
error of only $\approx$100~K would be sufficient to resolve the
discrepancy with the data.  Therefore, while acquiring resolved
spectra of the two components could help refine the temperature
determination, the systematic uncertainties in the atmosphere models
will still hamper accurate placement on the H-R diagram.  We discuss
this further in the next section.

The opposite interpretation is that the evolutionary models are
incorrect, leading to a $\approx$50\% overprediction of the
luminosities and a $\approx$20\% overprediction of the radii, given
the component masses.  Equivalently, if one were simply to assume the
H-R~diagram positions are accurate, the evolutionary models would
suggest individual masses of around 0.05~\Msun\ and 0.06~\Msun\ from
the Lyon and Tucson models, respectively, \ie, nearly a factor of two
overestimate in the masses.  While it may be that the evolutionary
models are so substantially incorrect, such a conclusion is not
compelling at this point, given the plausible errors in the \Teff\
determinations.\footnote{Note that a systematic error in our
  model-derived mass ratio cannot resolve the H-R diagram discrepancy.
  Since the ratio is basically unity, correcting any errors in $q$
  would move one evolutionary track closer to one component, while the
  other track would move farther away from the other component.  The
  evolutionary models could be brought into agreement with the data
  for component~A for $q\lesssim0.4$, but this is implausible given
  the nearly equal magnitudes of the two components.  Such a small $q$
  would also exacerbate the disagreement between the observations and
  the models for component~B.}

Direct mass determinations for the individual components from radial
velocity monitoring and/or absolute orbital astrometry will help to
further characterize the system.  Such data will directly test the
$q=0.936_{-0.008}^{+0.012}$ determined from the evolutionary tracks.  The expected
maximum radial velocity difference of the two components is only
4.6~\kms.  Since the two components are nearly equal mass and
brightness, the orbital motion will be very difficult to detect in the
integrated-light spectrum.  Resolved AO spectroscopy will be required,
and the small amplitude will make it a challenging measurement given
the few~\kms\ accuracy that has been achieved for T~dwarfs on the
largest existing telescopes \citep{2007ApJ...666.1205Z}.

Individual mass measurements can in principle also test the
evolutionary tracks directly.  One can estimate the age of each
component from its mass and luminosity (as we have done using the
total mass) and see if the ages indicate coevality for the system.
However, given the near-equal flux ratio of this system (and most
ultracool binaries), this coevality test is unlikely to be very
discriminating.  Moreover, individual masses cannot resolve the
discrepancy seen in the H-R diagram (Figure~\ref{fig:hrd-masses}),
which largely arises from the uncertainties in the model atmospheres.

\subsection{Future Tests of Theory with Field Substellar
  Binaries \label{sec:tests}}

With the advent of LGS AO on the largest ground-based telescopes, we
can expect an increasing number of dynamical masses for ultracool
field dwarfs in the near-future.
The most useful systems for testing theory will be those with both
independent mass and age determinations, namely binaries that are
associated with open clusters/groups and/or field stars of known age.
The former will present a significant technical challenge, \eg,
ultracool binaries in the Hyades ($d=46.3\pm0.3$~pc;
\citealp{1998A&A...331...81P}) and Pleiades ($d=133.5\pm1.2$~pc;
\citealp{2005AJ....129.1616S}) with suitably short orbital periods are
unresolvable with current technology and thus none are currently
known.  Ultracool binary companions to field stars are extremely rare
and thus while very valuable systems, these will only probe a very
limited range of spectral type, age, and mass: only four systems are
known with suitably short orbital periods ($P\lesssim50$~yr) --- the
T1+T6 binary \eInd~Bab \citep{2004A&A...413.1029M}, the L4+L4 binary
HD~130948BC \citep{2002ApJ...567L.133P}, the L4.5+L6 binary GJ~417BC
\citep{2003AJ....126.1526B, 2003AJ....125.3302G}, and the L4.5+L4.5
binary GJ~1001BC \citep{2004AJ....128.1733G}.
Therefore, there is significant motivation to develop analyses that
employ masses derived from the much more numerous field binaries.
In this regard and as illustrated by our analysis for \twomassbin, one
can identify two orthogonal pathways to confront theory:
(1)~comparison to evolutionary tracks and (2)~comparison to
atmospheric models.

\subsubsection{Comparison to evolutionary tracks (``H-R Diagram
  Test'')}
Direct measurements of \Lbol, \Teff, and mass (or age) for brown
dwarfs enable use of the H-R~diagram, by comparing the observations to
evolutionary tracks that correspond to the measured masses of the
objects.  As illustrated by Figure~\ref{fig:hrd-masses}, the Lyon and
Tucson tracks differ at the 5--10\% level in mass, and thus mass
determinations of 2--3\% accuracy could discriminate between the two
models, if \Lbol\ and \Teff\ can be well-measured.  (Improvements in
the parallaxes of many ultracool binaries will also be needed to
achieve such accurate masses.)
Accurate measurements for \Lbol\ are largely straight-forward, as good
as a few percent (\eg, \S~\ref{sec:lbol}).  However, direct \Teff\
determinations are extremely challenging, since radius measurements
are needed.  Brown dwarfs are too small and faint to be resolved with
current or planned interferometers, and no eclipsing ultracool field
binaries are yet known.  Thus, \Teff\ must be derived from modeling
the observed colors, magnitudes, and/or spectra; the approach
currently suffers from uncertainties at the level of a few to several
hundred Kelvin and systematic errors that are difficult to quantify
\citep[e.g.][]{2007arXiv0711.0801C}.  In comparison,
Figure~\ref{fig:hrd-masses} shows that \Teff\ determinations good to
$\lesssim$30~K are needed.  Therefore, decisive tests of evolutionary
tracks using field binaries will be challenged by this uncertainty in
\Teff, in the absence of direct radius measurements.

\subsubsection{Comparison to atmospheric models (``Age/Mass Benchmark Test'')}
Brown dwarfs obey a mass-luminosity-age ($M, \Lbol, t$) relation, and
for most field objects neither the mass nor the age is known.  A
commonly used approach to circumvent this limitation is to study brown
dwarfs that are companions to main-sequence stars, where (indirect)
age estimates are available from the primary star
\citep[e.g.][]{2000ApJ...541..374S, 2001ApJ...556..373G,
  2004A&A...413.1029M, 2006astro.ph..7514M, 2006liu-hd3651b,
  2006astro.ph..9556B}.  This approach can also be applied to members
of coeval clusters/groups and companions to post-main-sequence stars
of known age \citep[e.g.][]{1999ApJ...519..834K, 2006MNRAS.368.1281P}.
In these situations, \Lbol\ and $t$ are known, and combined with
evolutionary models, one can derive $M$ and consequently \Teff\ and
\logg.  Then the observed colors, magnitudes, and spectra can test the
accuracy of atmospheric models with the same \Teff\ and \logg.
Examination of the known ``age benchmark'' T dwarfs in this fashion
finds that the properties deduced from atmospheric models are in good
agreement with those from the evolutionary models, within the
uncertainties in the ages and metallicities of the primary stars
\citep{2006astro.ph..9556B, 2007arXiv0705.2602L, 2008arXiv0804.1386L}.

We suggest that, in an analogous fashion, field binaries with known
masses can also serve as ``benchmark'' objects.  In this case, $M$ and
\Lbol\ are known, and combined with evolutionary models, one can
derive $t$, as we have done in \S~4.1.  This provides \Teff\ and
\logg\ and thereby allows tests of atmospheric models.  The chain of
analysis is identical to brown dwarf companions of known age: given
independent knowledge of two quantities out of $\{M, \Lbol, t\}$, use
evolutionary models to derive the third.\footnote{The analogy between
  mass benchmarks and age benchmarks is an imperfect one, since brown
  dwarf companions also have metallicity determinations from their
  parent star, whereas field binaries do not.  However, current
  studies of age benchmarks largely rely on evolutionary models
  computed for solar metallicity.
} In both cases, these benchmarks can also serve as anchor points for
direct empirical calibration of spectroscopic diagnostics of \logg\
and \Teff\ \citep[e.g.][]{2006MNRAS.368.1281P, 2006ApJ...639.1095B}.

At face value, using objects that are age benchmarks or mass
benchmarks is less fundamental than direct tests of the evolutionary
models using the H-R diagram.  However, in practice the Benchmark Test
is much more feasible to implement and subject to much smaller
systematic errors.  In the absence of direct radius measurements, the
H-R Diagram Test is held hostage to the systematic errors in the
determination of \Teff\ from atmospheric models.  In contrast, the
Benchmark Test relies on the evolutionary models, which are thought to
be more robust \citep[e.g.][]{2000ApJ...542..464C}.  In short, given
the choice of relying on atmospheric models (H-R Diagram Test) or
evolutionary models (Benchmark Test), the evolutionary models are
likely to be preferred.

To assess the relative utility of ``age benchmarks'' (brown dwarf
companions to stars) compared to ``mass benchmarks'' (brown dwarfs
with dynamical masses), we turn to Equation~\ref{eqn:lbol}.  For an
object with a measured \Lbol\ and ignoring the weak dependence on
$\kappa_R$, given a measurement of $M$ or $t$ with accompanying
uncertainty of $\delta{M}$ or $\delta{t}$, the fractional error in the
remaining quantity is related by:
\begin{equation}
  \frac{\delta{t}}{t} = 2.03\ \frac{\delta{M}}{M}\ .
\end{equation}
Typical uncertainties in the ages of main-sequence field stars are
about 50--100\% \citep[e.g.][]{2001AJ....121.3235K, 2006liu-hd3651b,
  2006astro.ph..7514M}, and thus age-benchmark objects would have a
25--50\% uncertainty in the mass inferred from evolutionary
models.\footnote{Barnes (2007) find that ages for solar-type stars
  derived from gyrochronology can have errors of only 15--20\%.}
In contrast, dynamical masses of $\approx$5--10\% accuracy will be
possible in the next few years for many ultracool dwarfs, as we have
already achieved with \twomassbin, and hence mass-benchmark objects
will have only a $\approx$10--20\% error in the age inferred from
evolutionary models.  So overall, age benchmarks can be expected to
have $\{\delta{t}/t, \delta{M}/M\} \approx$~\{50--100\%, 25--50\%\}
while mass benchmarks would have uncertainties of order
$\{\delta{t}/t, \delta{M}/M\} \approx$~\{10--20\%, 5--10\%\}.

We can use the analytic fits to evolutionary models from \citet{bur01}
to gauge the relative accuracy on \Teff\ and \logg\ derived from both
types of benchmarks.  Using standard error propagation and assuming
uncorrelated errors, we find for \underline{mass benchmarks}:
\begin{eqnarray}
\frac{\delta{\Teff}}{\Teff} & = & 
   \sqrt{\left(0.180 \frac{\delta{M}}{M}\right)^2 + 
         \left(0.246 \frac{\delta{\Lbol}}{\Lbol}\right)^2} \\
\delta{\logg}               & = &
   \sqrt{\left(1.63 \frac{\delta{M}}{M}\right)^2 + 
         \left(0.088 \frac{\delta{\Lbol}}{\Lbol}\right)^2} 
\end{eqnarray}
\begin{eqnarray}
\delta\Teff & = & 9~{\rm K} \left(\frac{\Teff}{1000~{\rm K}}\right) \ 
                     \sqrt{\left(\frac{\delta{M}/{M}}{0.05}\right)^2 + 
                           \left(2.73\ \frac{\delta{\Lbol}/{\Lbol}}{0.10}\right)^2}   \\
\delta\logg & = & 0.082~{\rm dex} \ \sqrt{\left(\frac{\delta{M}/{M}}{0.05}\right)^2 + 
                                      \left(1.08 \frac{\delta{\Lbol}/L}{0.10}\right)^2}
\end{eqnarray}
where $\delta{\Teff}, \delta{M}, \delta{\Lbol},$ and $\delta{\logg}$
are the uncertainties in the temperture, mass, luminosity, and surface
gravity, respectively.  And then for \underline{age benchmarks}, we
find:
\begin{eqnarray}
\frac{\delta{\Teff}}{\Teff} & = & 
   \sqrt{\left(0.089 \frac{\delta{t}}{t}\right)^2 + 
         \left(0.314 \frac{\delta{\Lbol}}{\Lbol}\right)^2} \\
\delta{\logg}               & = &
   \sqrt{\left(0.734 \frac{\delta{t}}{t}\right)^2 + 
         \left(0.517 \frac{\delta{\Lbol}}{\Lbol}\right)^2} 
\end{eqnarray}
\begin{eqnarray}
\delta\Teff & = & 44~{\rm K} \left(\frac{\Teff}{1000~{\rm K}}\right) \ 
                     \sqrt{\left(\frac{\delta{t}/{t}}{0.5}\right)^2 + 
                           \left(0.71\ \frac{\delta{\Lbol}/{\Lbol}}{0.10}\right)^2}   \\
\delta\logg & = & 0.37~{\rm dex} \ \sqrt{\left(\frac{\delta{t}/{t}}{0.5}\right)^2 + 
                                      \left(0.14 \frac{\delta{\Lbol}/\Lbol}{0.10}\right)^2}.
\end{eqnarray}
Thus with representative values for the fractional errors in age
(50\%), luminosity(10\%), and mass (5\%), we see that \Teff\ and
\logg\ are better constrained by a factor of $\approx$5 using mass
benchmarks than age benchmarks.  Figure~\ref{fig:benchmarks} plots the
derived analytic estimates for both types of benchmarks.  These
contour plots provide a convenient means to gauge the expected errors
in \Teff\ and \logg\ determinations from benchmarks.  The morphology
of the contours also illustrates whether the observational errors in
the age, mass, and/or \Lbol\ dominate the uncertainties in \Teff\ and
\logg.  For the specific case of \twomassbin, there is good agreement
between the analytic estimates and the values derived directly from
the actual evolutionary models (Table~\ref{table:evolmodels}).


\section{Conclusions}

We have determined the first dynamical mass for a binary T~dwarf, the
T5.0+T5.5 system \twomassbin, by combining six epochs of Keck LGS AO
imaging from 2005--2008 with three epochs of \HST\ imaging obtained in
2000 and 2006.  Both datasets achieve milliarcsecond accuracy or
better for the relative astrometry of the two components and are
validated through extensive testing with images of simulated binaries.
We employ a Markov Chain Monte Carlo analysis to determine the orbital
parameters and their uncertainties.  The time baseline of our complete
dataset covers about half of the total period.  We find that the
orbital motion of the binary is viewed in an almost edge-on
orientation and has a modest eccentricity.  
Our determination of a $15.1_{-1.6}^{+2.3}$~yr orbital period is
significantly longer than the original 4-year estimate, as by chance
the binary was at a very small projected separation when discovered in
2000.

The total mass of the system is $0.056\pm0.003~\Msun$
($59\pm4$~\Mjup), including the uncertainty in the parallax.  This is
the second brown dwarf binary directly confirmed, the first among the
field population.  It is also the coolest and lowest mass binary with
a dynamical mass determination to date.  

With very accurate measurements of the total mass and the bolometric
luminosity ($\log(\Lbol/\Lsun) = -4.751\pm0.011$), we use the Tucson
and Lyon evolutionary tracks to determine the remaining physical
properties for the system.  The two sets of models give largely
consistent results, which highlights the difficult of distinguishing
between them even with such precise observational data.  We average
the model results to represent the final determinations.  We find a
relatively youthful age for the system of 0.79$\pm$0.09~Gyr
(1$\sigma$), consistent with its low tangential velocity relative to
other field T~dwarfs.
The remaining physical parameters of the individual components are
then fully determined: radii of \hbox{$0.0986\pm0.0015~\Rsun$} and
\hbox{$0.0993\pm0.0017~\Rsun$}, effective temperatures of $1028 \pm
17$~K and $978 \pm 17$~K, surface gravities of \hbox{$\logg = 4.91 \pm
  0.04$} and $4.87 \pm 0.04$, and masses of $0.0287\pm0.0016~\Msun$
($30.1\pm1.7~\Mjup$) and $0.0269\pm0.0016~\Msun$ ($28.2\pm1.7~\Mjup$)
for components~A and~B, respectively.  We take care to account for the
covariances inherent in the measurement uncertainties, by using a
Monte Carlo approach to derive these physical quantities from the
evolutionary models.  Our approach also assumes that the system is
coeval and composed of only two components.

These precise determinations for 2MASS~J1534$-$2952A and~B are in
general accord with the \Teff\ and \logg\ values found previously for
field T~dwarfs based on model atmospheres and with the ages of
T~dwarfs predicted by Monte Carlo simulations of the solar
neighborhood.  However, upon closer scrutiny, there are two potential
discrepancies with past studies.  Both suggest that the temperatures
of field T~dwarfs may be overestimated by $\approx$100~K, though we
stress that the two discrepancies must arise from independent effects.
(1)~The temperatures of 2MASS~J1534$-$2952A and~B appear to be cooler
than field objects of comparable spectral type.  This resembles
discrepancies previously noted by \citet{2006astro.ph..7514M} and
\citet{2006astro.ph..9464L} for late-L/early-T dwarfs that are
companions to young main-sequence stars.  They have hypothesized that
the effect is due to the gravity sensitivity of the L/T transition.
The fact that this discrepancy also occurs for \twomassbin\ suggests
instead that the problem may arise from a factor of $\approx6\pm3$
overestimate in the adopted ages of field objects when determining
their temperatures using evolutionary tracks.  Ages of
$\sim$0.3--1.0~Gyr are preferred based on this binary.  (2) The
temperatures of 2MASS~J1534$-$2952A and~B are slightly cooler than
inferred for other mid-T~dwarfs from model atmospheres.  Detailed
analysis of the system's integrated-light and resolved spectra with
model atmospheres is needed to directly assess the \Teff\ and \logg\
of the two components and to refine the comparison with the values
derived from evolutionary models.

The positions of the two components on the H-R diagram are discrepant
with theoretical evolutionary tracks corresponding to their individual
masses.  In fact, taken at face value, using the H-R diagram positions
to infer masses from the evolutionary tracks would lead to masses of
$\approx$0.05--0.06~\Msun, about a factor of two larger than the
actual measured masses.  While this discrepancy could stem from large
systematic errors in the luminosities ($\sim$50\% errors) and/or radii
($\sim$20\% errors) predicted by evolutionary models, the likely cause
is that temperatures from model atmospheres are too warm by
$\approx$100~K for mid-T dwarfs.  This highlights the need for
continued improvements to the model atmospheres.

Future monitoring of \twomassbin\ will help to refine its orbit and
its dynamical mass.  The orbital separation of the system is now
rapidly decreasing and will not be readily resolvable again until
around 2011.  At the same time, an improved parallax for the system
will be required --- the uncertainty in the total mass from the orbit
fitting is 3\%, compared to the 5\% that arises from the uncertainty
in the parallax.  Radial velocity monitoring and/or absolute
astrometry will directly determine the individual masses and test if
the system is a higher order multiple.  However, given the very
similar fluxes of the two components (implying nearly equal mass),
individual mass measurements are unlikely to resolve the discordant
H-R diagram position of the two components relative to evolutionary
tracks.  This problem is likely driven by the systematic uncertainties
in current model atmospheres for T~dwarfs.

The fundamental characteristic of the field population is that it
spans a range of (largely unknown) ages.  However despite this
uncertainty, field brown dwarf binaries can strongly test theoretical
models, if analyzed appropriately.  These systems will be especially
valuable in light of the current paucity of eclipsing field ultracool
binaries and resolvable, short-period ultracool binaries in open
clusters/groups.\footnote{For reference, about 1~out of a~1000 stars
  are eclipsing binaries, whereas there are only $\sim$600~L~and
  T~dwarfs known.}  Specifically, attempts to directly test different
evolutionary tracks by placing ultracool objects on the H-R diagram
(the ``H-R Diagram Test'') will be challenging, given the similarity
between the tracks and the difficulty in independently determining
\Teff\ with model atmospheres.  Instead, atmosphere models can be
confronted against \logg\ and \Teff\ values for ultracool objects as
derived from the evolutionary models, which can be exceptionally
precise (the ``Benchmark Test'').  This approach has previously been
applied to single brown dwarfs that are companions to stars of known
age (``age benchmarks'').  We suggest that in an analogous fashion,
field ultracool binaries with dynamical mass determinations (``mass
benchmarks'') can test the model atmospheres.  In fact, given the
plausible observational uncertainties, mass benchmarks are likely to
provide stronger constraints (by a factor of $\approx$5) on \logg\ and
\Teff\ than age benchmarks, since dynamical masses can be determined
far more accurately than ages for main-sequence field stars.  With the
widespread advent of LGS AO on the largest ground-based telescopes, we
can look forward to a rapid increase in dynamical mass determinations
for low-mass field binaries and thus substantial advances in our
understanding of the properties and evolution of substellar objects.


\acknowledgments

We gratefully acknowledge the Keck LGS AO team for their exceptional
efforts in bringing the LGS AO system to fruition.  It is a pleasure
to thank Antonin Bouchez, David LeMignant, Marcos van Dam, Randy
Campbell, Al Conrad, Jim Lyke, Hien Tran, Robert LaFon, Kenny Graves,
Cindy Wilburn, Joel Aycock, Terry Stickel, Gary Punawai, and the Keck
Observatory staff for assistance with the observations.  We thank Alan
Stockton for a fortuitous swap of observing nights in Spring 2005;
Brian Cameron for sharing his NIRC2 instrumental distortion analysis;
Hai Fu for IDL plotting assistance; Michael Cushing and Mark Pitts for
asssistance with the IRTF/SpeX spectroscopy; Adam Burrows and Isabelle
Baraffe for finely gridded evolutionary tracks; and Thierry Forveille
and Adam Burgasser for careful readings of the manuscript.  We have
benefitted from enlightening discussions with Michael Cushing, Adam
Burrows, and Isabelle Baraffe about substellar models; Thierry
Forveille about low-mass binaries; Brian Cameron about astrometry with
NIRC2; and Jay Anderson about astrometry with \HST.
Our research has employed the 2MASS data products; NASA's
Astrophysical Data System; the SIMBAD database operated at CDS,
Strasbourg, France; the M, L, and T~dwarf compendium housed at
DwarfArchives.org and maintained by Chris Gelino, Davy Kirkpatrick,
and Adam Burgasser \citep{2003IAUS..211..189K, 2004AAS...205.1113G};
and the SpeX Prism Spectral Libraries maintained by Adam Burgasser at
http://www.browndwarfs.org/spexprism.
MCL and TJD acknowledge support for this work from NSF grant
AST-0507833 and an Alfred P. Sloan Research Fellowship.  MJI
acknowledges Michelson Fellowship support from the Michelson Science
Center and the NASA Navigator Program.
Finally, the authors wish to recognize and acknowledge the very
significant cultural role and reverence that the summit of Mauna Kea has
always had within the indigenous Hawaiian community.  We are most
fortunate to have the opportunity to conduct observations from this
mountain.

{\it Facilities:} \facility{Keck II Telescope (LGS AO, NIRC2)},
\facility{\HST\ (WFPC2, ACS)}, \facility{IRTF (SpeX)}

\clearpage



\begin{figure}
\vskip -1in
\hskip -0.3in
\centerline{\includegraphics[height=7.5in,angle=90]{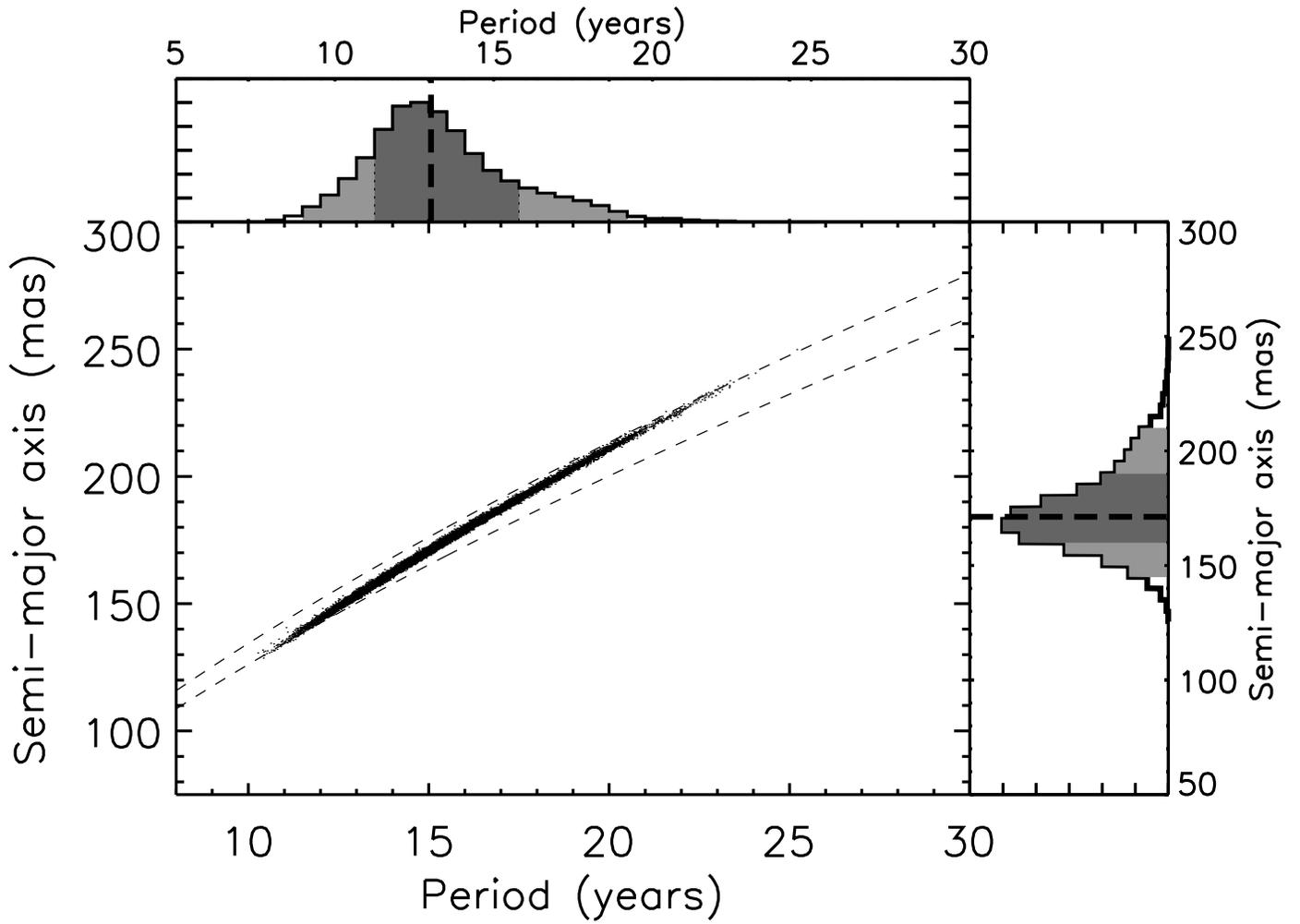}}
\vskip -2ex
\caption{\normalsize Results from MCMC determination of orbital period
  and semi-major axis.  See Figure~\ref{fig:orbit-p_e} for further
  explanation.  The dashed lines represent the $\pm$3$\sigma$
  confidence intervals of the mass probability
  distribution. \label{fig:orbit-p_a}}
\end{figure}

\begin{figure}
\vskip -1in
\centerline{\includegraphics[height=7.5in,angle=90]{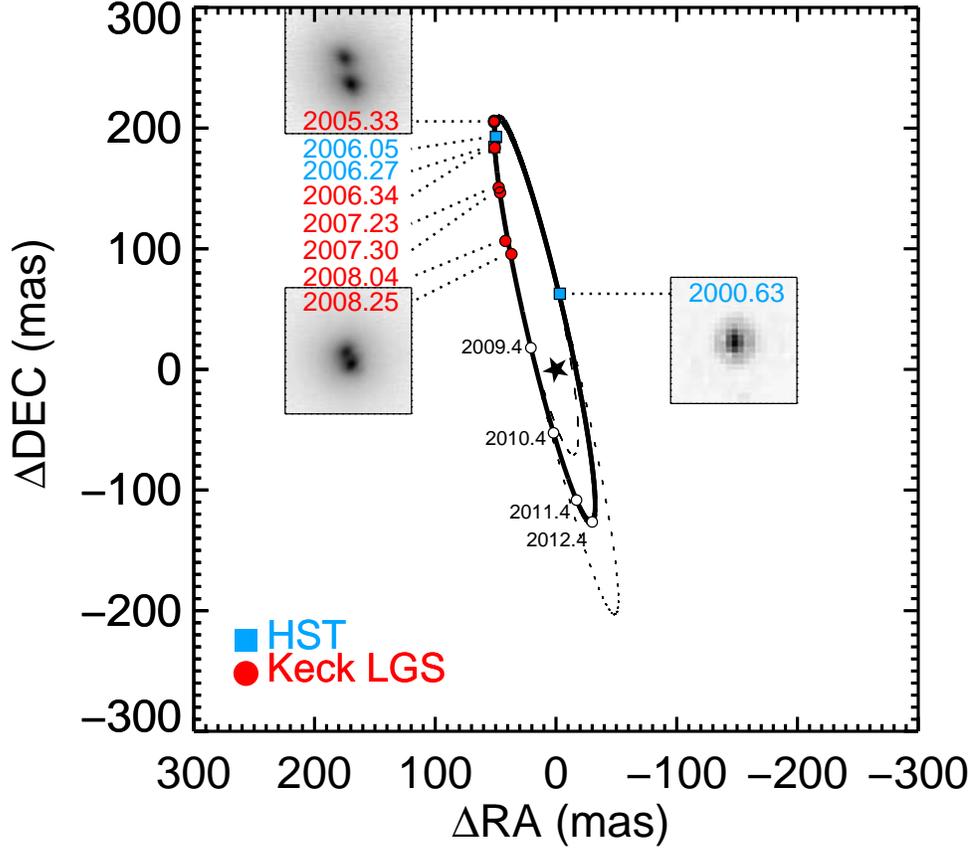}}
\vskip -2ex
\caption{\normalsize Keck LGS (red dot) and \HST\ (blue square)
  measurements of \twomassbin\ along with three representative orbits
  from the MCMC fitting.  Measurement errors are comparable to or
  smaller than the colored plotting symbols.  The solid line shows the
  best-fitting orbit with a 15.2-year period, and the two dashed
  orbits represent alternative long-period (20 year) and short-period
  (12 year) solutions, chosen to represent the plausible range in
  period.  The three orbits have reduced chi-square values of 0.9,
  1.0, and 1.1.  The empty circles are the location of the secondary
  in future years as predicted by the 15.2-year period orbit.  The
  image insets are 1\arcsec\ on a side, displayed with a square-root
  stretch.  (The HST/WFPC2 image cutout has been rotated so that North
  is up.  See Figure~\ref{fig:images} for the most accurate
  representation of the original data.)
  \label{fig:orbit-orbit}}
\end{figure}

\begin{figure}
\vskip -1in
\hskip -0.8in
\includegraphics[height=5in,angle=90]{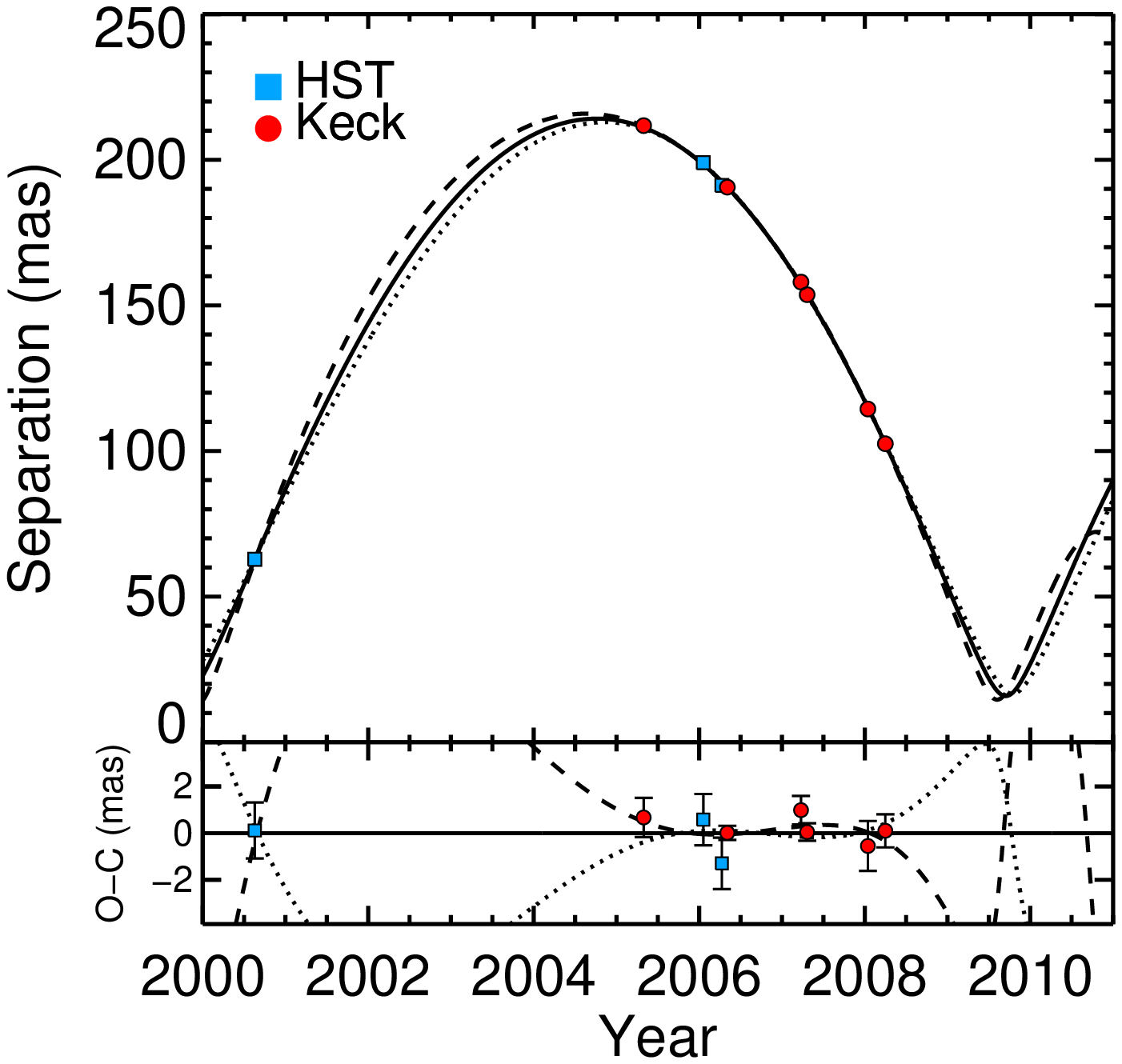}
\hskip -1.5in
\includegraphics[height=5in,angle=90]{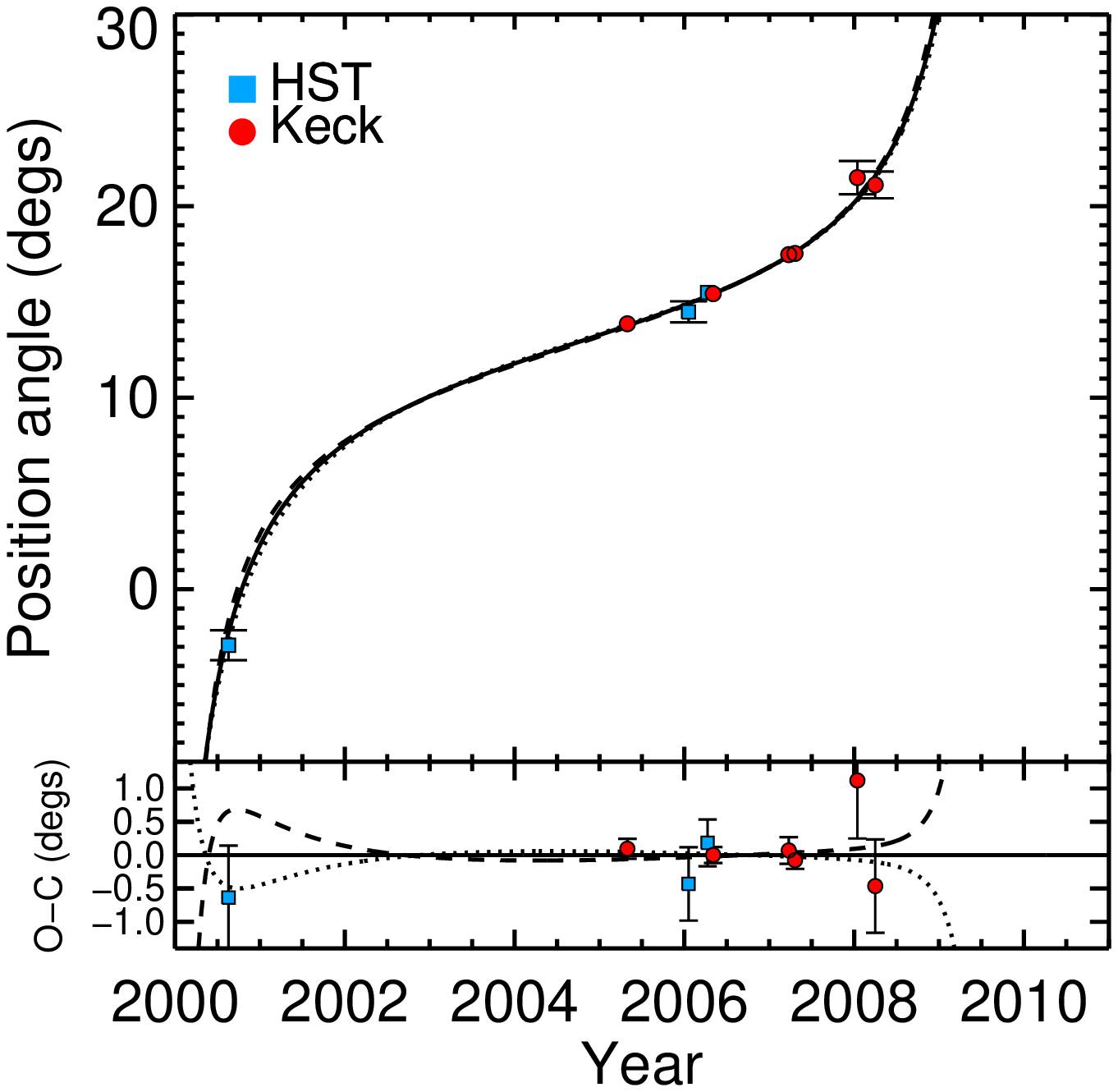}
\vskip -2ex
\caption{\normalsize Keck LGS and \HST\ measurements of \twomassbin's
  separation ({\em left}) and PA ({\em right}), along with the three
  orbits plotted in Figure~\ref{fig:orbit-orbit}.  The measurement
  errors are comparable to or smaller than the plotting symbols,
  except for the three data points shown in the PA plot.  The bottom
  panel of each plot shows the difference of the observed astrometry
  and the best-fitting (15.2-year period)
  orbit. \label{fig:orbit-sep}}
\end{figure}

\begin{figure}
\vskip -1in
\centerline{\includegraphics[height=7.5in,angle=90]{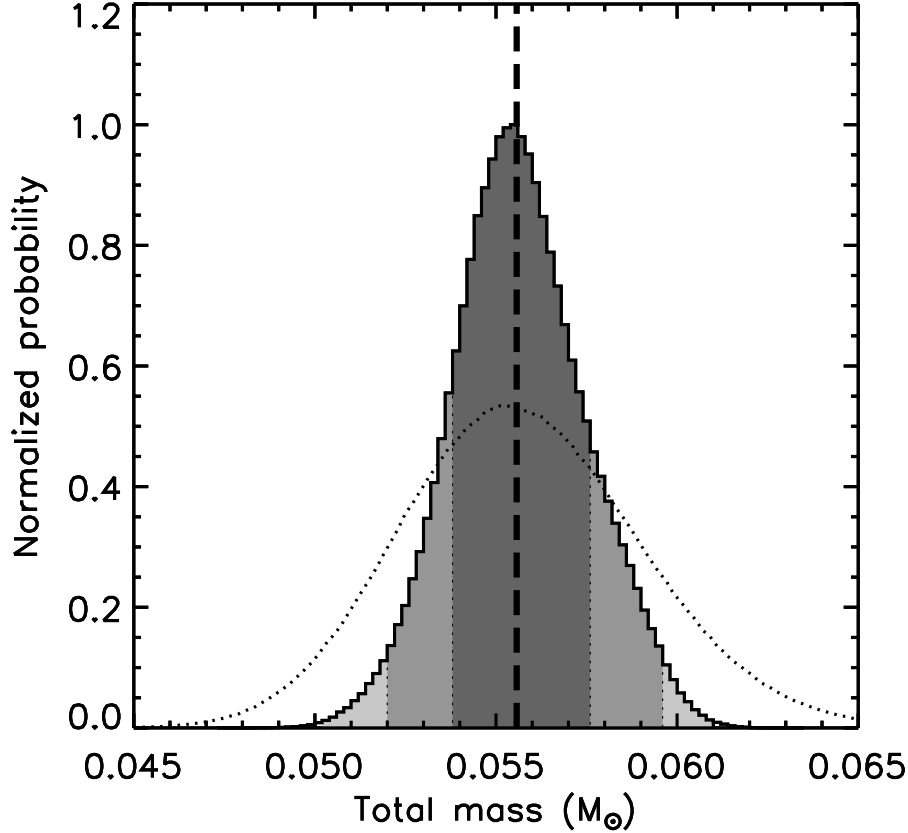}}
\vskip 2ex
\caption{\normalsize Total mass distribution from our MCMC analysis.
  The histogram is shaded to indicate the derived 68.3\%, 95.5\%, and
  99.7\% confidence regions, which correspond to 1$\sigma$, 2$\sigma$,
  and 3$\sigma$ for a normal distribution. The solid vertical line
  represents the median value of 0.0556~\Msun. The standard deviation
  of the distribution is 0.0018~\Msun. The wider, unshaded histogram
  shows the final mass distribution, after accounting for the
  additional 4.9\% error due primarily to the uncertainty in the
  binary's parallax --- the result is essentially gaussian with a
  standard deviation of 0.003~\Msun.  The confidence limits for both
  distributions are given in
  Table~\ref{table:orbits}.\label{fig:orbit-masses}}
\end{figure}

\begin{figure}
\vskip -1in
\centerline{\includegraphics[height=7.5in,angle=90]{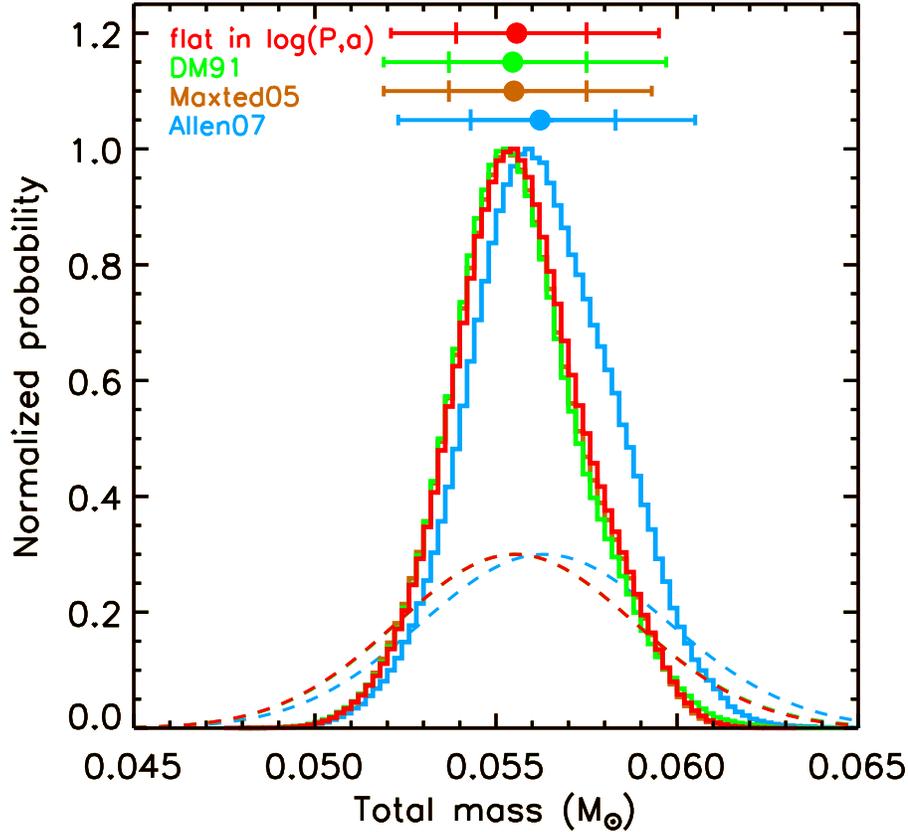}}
\vskip -4ex
\caption{\normalsize Total mass distribution from the MCMC analysis
  using four different priors: a distribution that is flat in the
  logarithm of the period and semi-major axis (our default
  assumption); the log-normal period distribution for solar-type stars
  from \citet{1991A&A...248..485D}; the log-normal semi-major
  distribution for field ultracool binaries from
  \citet{2007ApJ...668..492A}; and the truncated log-normal semi-major
  axis distribution from \citet{2005MNRAS.362L..45M}.  The
  median value is indicated by the filled circle and the two sets of
  errors bars indicate the 68.3\% and 95.5\% confidence regions, which
  correspond to 1$\sigma$ and 2$\sigma$ for a normal distribution.
  The dashed curves at the bottom show the resulting mass
  distributions after accounting for the uncertainty in the parallax.
  While the \citet{2007ApJ...668..492A} prior favors slightly higher
  masses, all four priors give very consistent
  results. \label{fig:priors}}
\end{figure}

\begin{figure}
\vskip -1in
\hskip -0.1in
\centerline{\includegraphics[height=7.5in,angle=90]{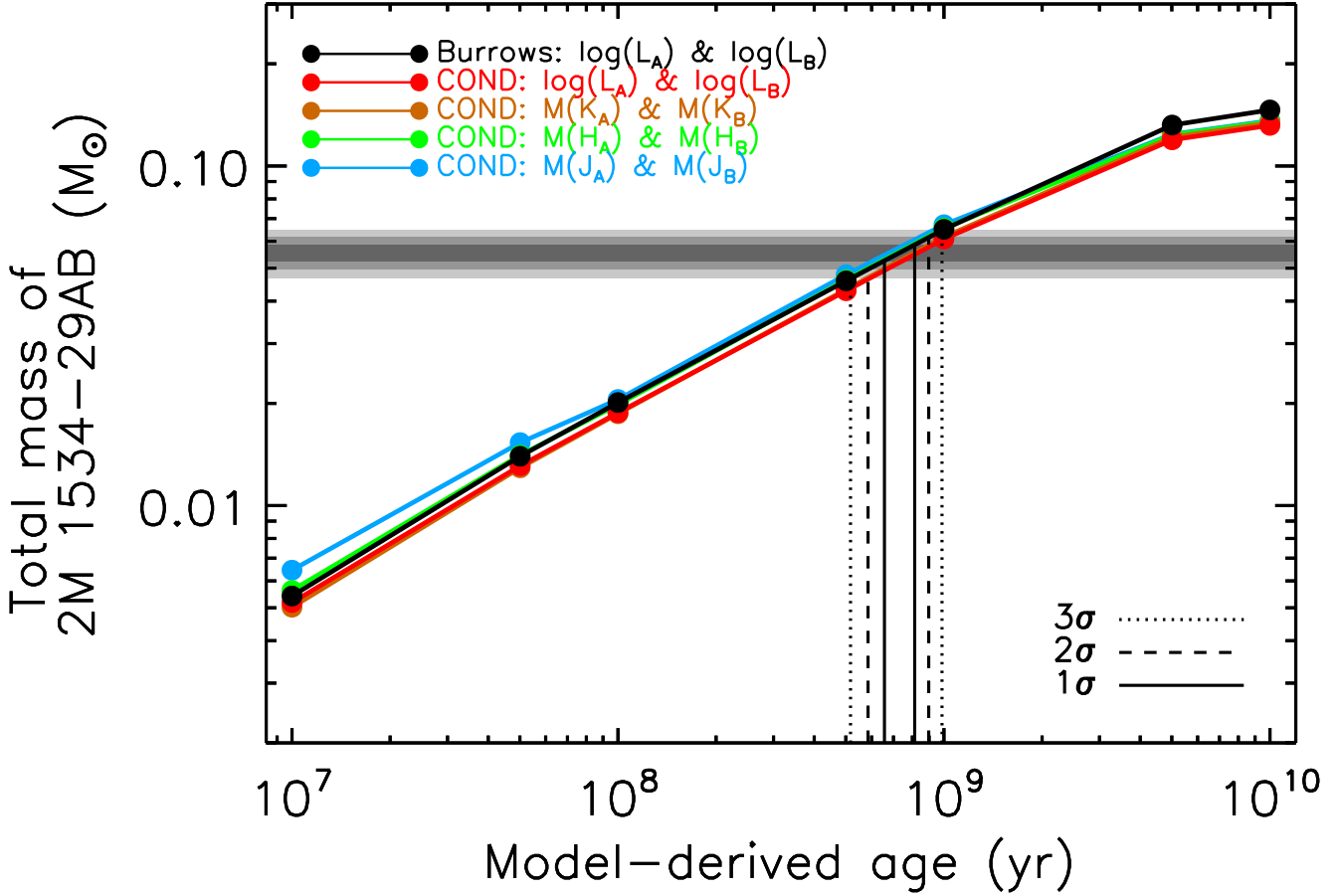}}
\vskip -4ex
\caption{\normalsize Determination of the age of the \twomassbin\
  system from evolutionary models using the observed
  magnitudes/luminosity and total mass.  Two sets of models are
  employed: the models from \citet{1997ApJ...491..856B} and the COND
  models from \citet{2003A&A...402..701B}.  The horizontal grey bars
  indicate our measured 1,~2, and~3$\sigma$ constraints on the total
  mass, and the vertical lines show the corresponding ages derived
  from the Burrows \etal\ models.  Both sets of models produce
  consistent results, given in Table~\ref{table:evolmodels}.  The COND
  models provide predictions for both the absolute magnitudes and
  bolometric luminosity, so all of these are shown.  (The two sets of
  models predict very similar \Lbol\ results, so the red and black
  model lines are indistinguishable on this plot.)  Note that these
  plotted curves are computed for an object with the fluxes of
  \twomassbin\ and are not generally applicable to other binaries.
  See \S~\ref{sec:ages} for details.  \label{fig:models-totalmass}}
\end{figure}

\begin{figure}
\vskip -2in
\hskip 0.3in
\includegraphics[height=5.5in,angle=90]{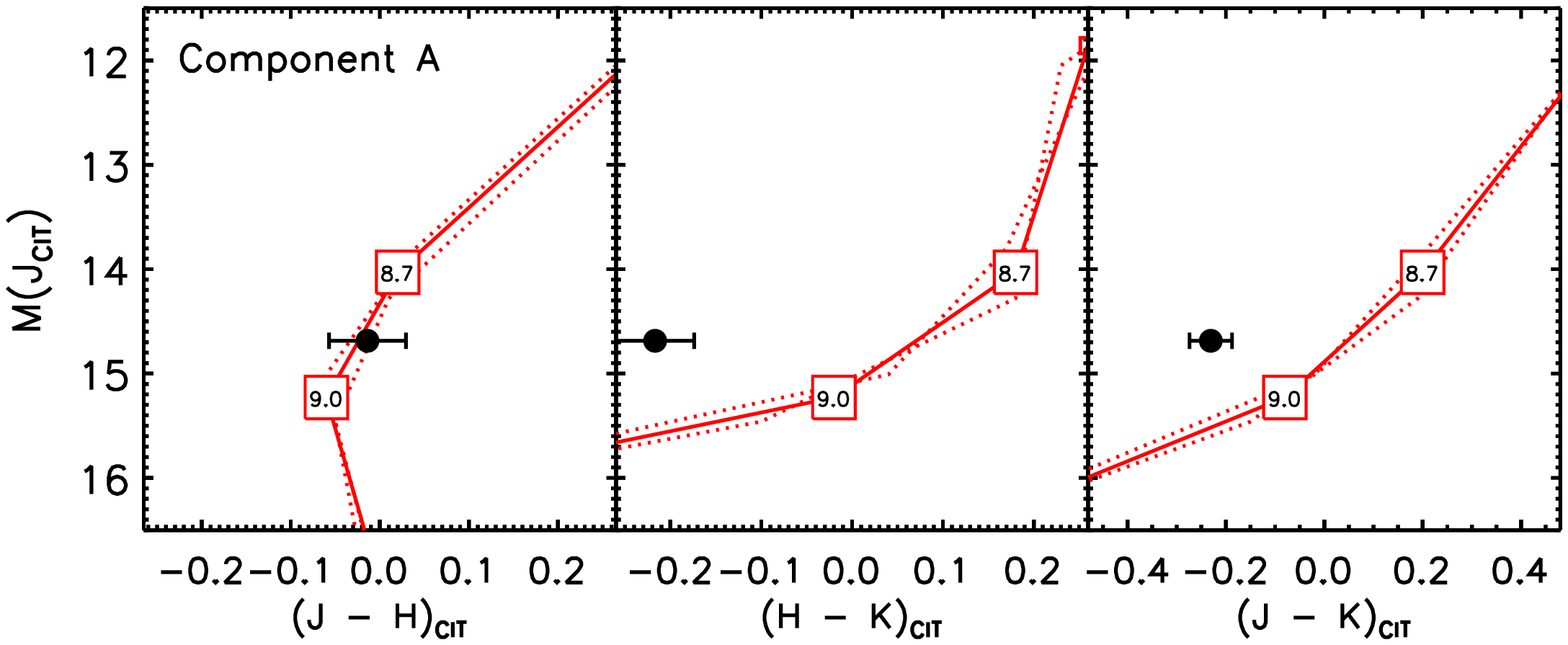}
\vskip -1.5in
\hskip 0.3in
\includegraphics[height=5.5in,angle=90]{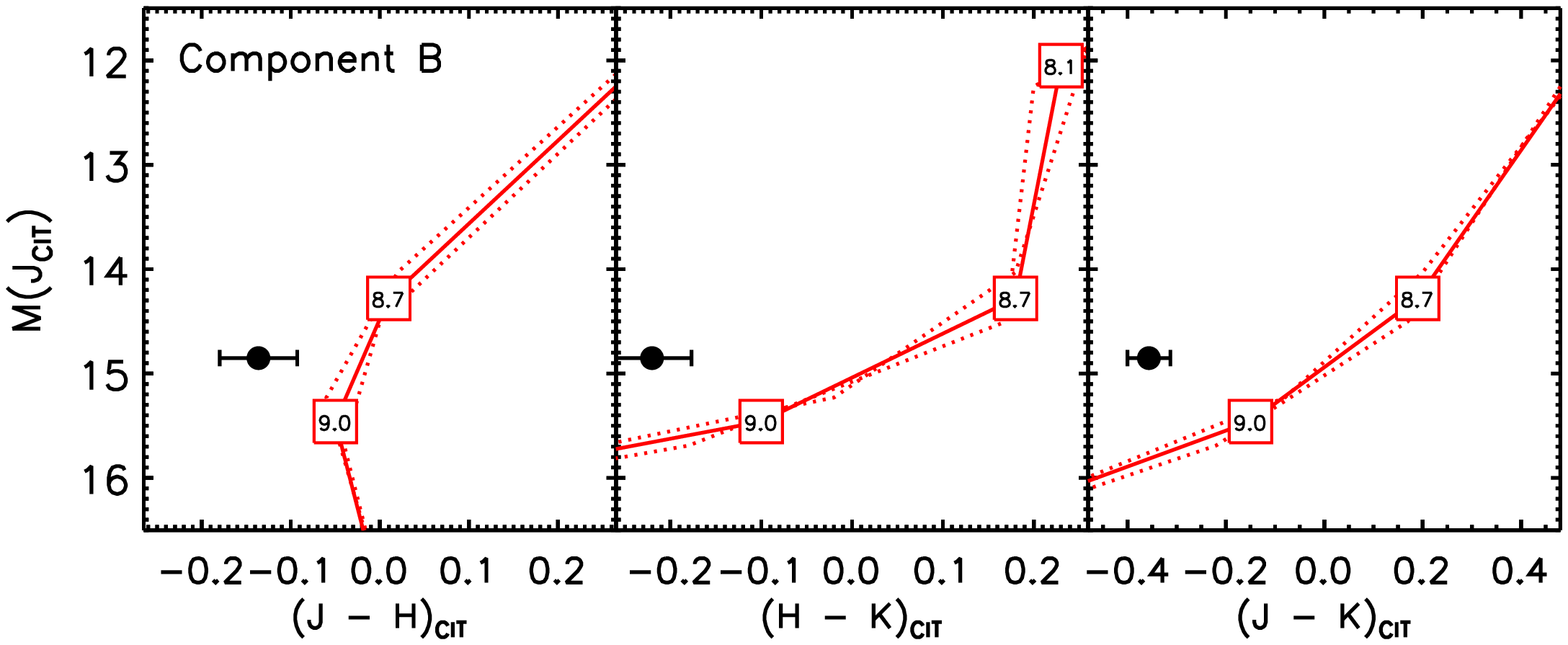}
\vskip -0.7in
\caption{\normalsize Color-magnitude diagrams for the individual
  components of \twomassbin\ compared to the Lyon COND models
  \citep{2003A&A...402..701B}, on the CIT photometric system.  The
  models corresponding to the individual component masses and their
  $\pm$1$\sigma$ range are plotted as solid lines and dotted lines,
  respectively.  The numbered boxes indicate the logarithm of the
  model age in Gyr.  The error bars on the absolute $J$-band magnitude
  is smaller than the plotting symbol.  The models do not match the
  data very well, which can be ascribed to the incomplete opacities in
  the model atmospheres.  \label{fig:cmd}}
\end{figure}

\begin{figure}
\vskip -1in
\centerline{\includegraphics[height=7in,angle=90]{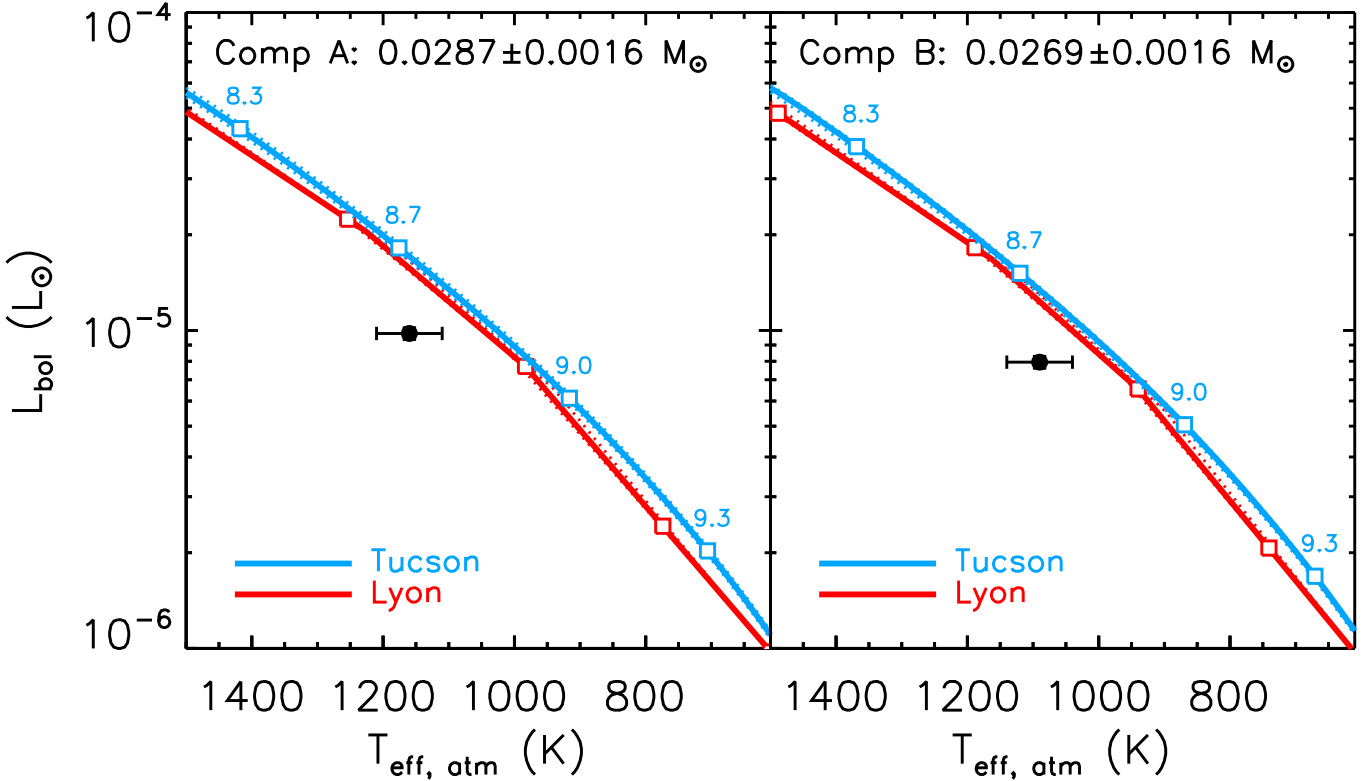}}
\vskip -2ex
\caption{\normalsize H-R diagram showing the individual components of
  \twomassbin\ compared to evolutionary tracks from the Tucson and
  Lyon groups.  The plotted symbols represent the data, with the
  effective temperatures determined from the spectral types and model
  atmosphere studies of field T~dwarfs (Equation~\ref{eqn:teff-atm}).
  The uncertainity in the \Lbol\ measurements are comparable to the
  vertical extent of the plotting symbols.  The solid line represents
  the median mass values determined for the individual components, and
  the dotted lines show the $\pm1\sigma$ ranges.  The embedded small
  squares demarcate logarithms of a set of ages; the ages for the
  Burrows models are labelled, and the (unlabeled) ages for the Lyon
  models have very similar positions along the plotted line.  There is
  a modest systematic disagreement between the evolutionary tracks and
  observations.\label{fig:hrd-masses}}
\end{figure}

\begin{figure}
\hskip -0.3in
\includegraphics[height=4.5in,angle=0]{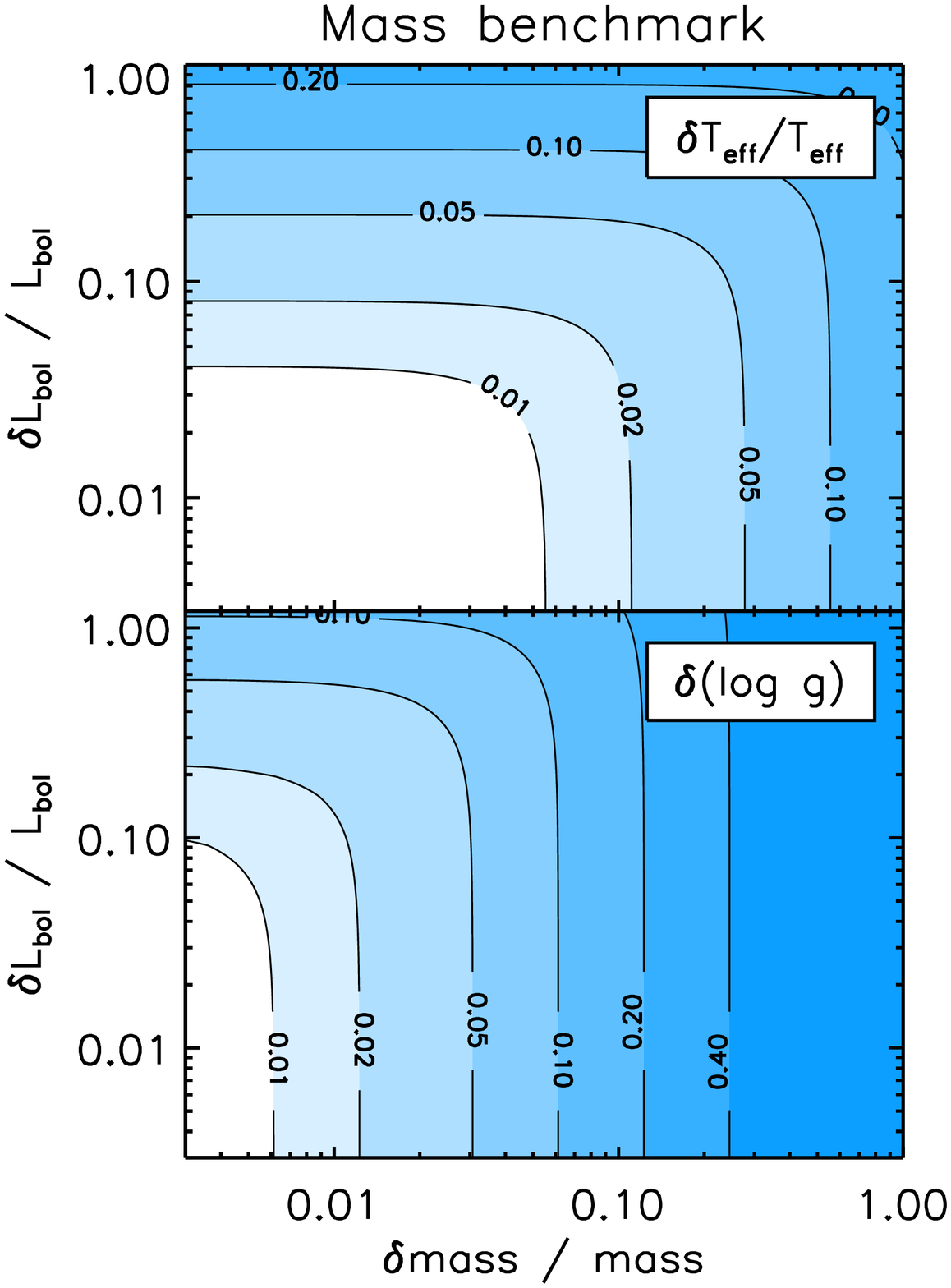}
\hskip 0.6in
\includegraphics[height=4.5in,angle=0]{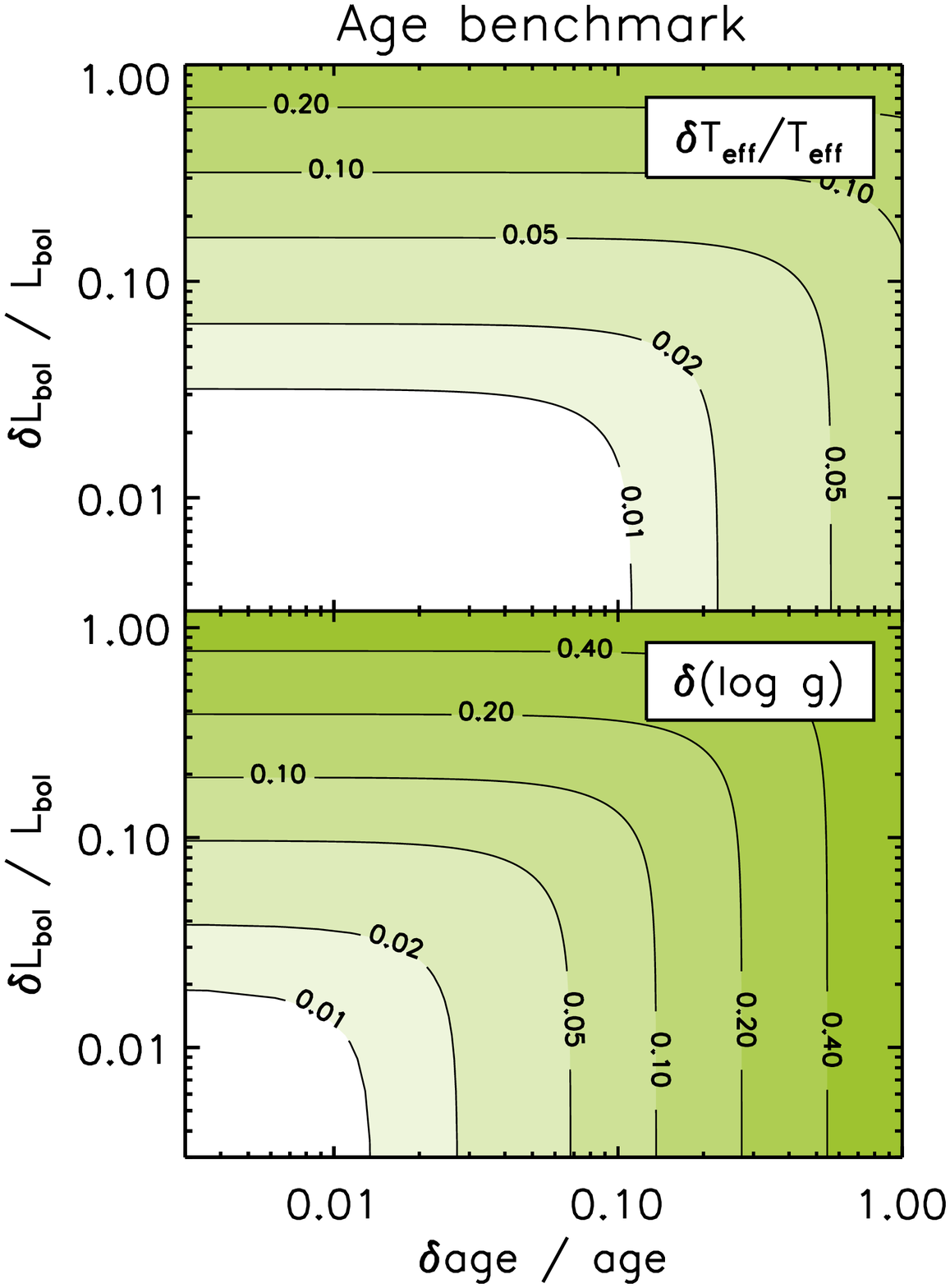}
\vskip 2ex
\caption{\normalsize Contour plots showing the uncertainties in
  determining \Teff\ and \logg\ using mass benchmarks (brown dwarfs
  with dynamical masses) and age benchmarks (brown dwarfs with known
  ages), derived from the \citet{bur01} analytic approximations to
  substellar evolution.  The x- and y-axes give the fractional error
  in the measurements of mass, age, and \Lbol, and the contours show
  the resulting fractional uncertainties in \Teff\ and \logg\ spaced
  from 1\% to 40\%. \label{fig:benchmarks}}
\end{figure}





\clearpage

\begin{deluxetable}{lccccccc}
\tablecaption{Keck LGS AO Observations \label{table:keck}}
\tabletypesize{\small}
\rotate
\tablewidth{0pt}
\tablehead{
  \colhead{Date} &
  \colhead{Filter\tablenotemark{a}} &
  \colhead{Airmass} &
  \colhead{FWHM} &
  \colhead{Strehl ratio} &
  \colhead{Separation\tablenotemark{b}} &
  \colhead{Position angle\tablenotemark{b}} &
  \colhead{$\Delta$mag} \\
  \colhead{(UT)} &
  \colhead{} &
  \colhead{} &
  \colhead{(mas)} &
  \colhead{} &
  \colhead{(mas)} &
  \colhead{(deg)} &
  \colhead{}
}
\startdata

2005-May-01 & $J$     & 1.66 & 102 $\pm$ 12 & 0.020 $\pm$ 0.002        & 211.3 $\pm$ 1.5 (1.5) & 14.1\phn\ $\pm$ 0.3\phn\ (0.3)\phn  & 0.163 $\pm$ 0.014  \\
            & $H$     & 1.63 &  86 $\pm$ 6  & 0.047 $\pm$ 0.006        & 211.7 $\pm$ 0.8 (0.8) & 13.86 $\pm$ 0.15 (0.13)             & 0.286 $\pm$ 0.011  \\
            & \Kp\    & 1.61 &  88 $\pm$ 6  & 0.101 $\pm$ 0.012        & 212.4 $\pm$ 1.1 (1.0) & 14.0\phn\ $\pm$ 0.2\phn\ (0.2)\phn  & 0.278 $\pm$ 0.021  \\

2006-May-05 & \Ks\    & 1.56 &  64 $\pm$ 3  & 0.210 $\pm$ 0.014        & 190.6 $\pm$ 0.3 (0.2) & 15.43 $\pm$ 0.12 (0.09)             & 0.282 $\pm$ 0.010  \\
2007-Mar-26 & $K$     & 1.56 &  82 $\pm$ 3  & 0.151 $\pm$ 0.016        & 158.0 $\pm$ 0.6 (0.6) & 17.5\phn\ $\pm$ 0.2\phn\ (0.19)     & 0.287 $\pm$ 0.012  \\
2007-Apr-22 & \Ks\    & 1.57 &  67 $\pm$ 5  & 0.20\phn\ $\pm$ 0.03\phn & 153.7 $\pm$ 0.4 (0.3) & 17.53 $\pm$ 0.13 (0.10)             & 0.269 $\pm$ 0.010  \\
2008-Jan-15 & \Ks\    & 2.05 & 100 $\pm$ 3  & 0.074 $\pm$ 0.002        & 114.4 $\pm$ 1.1 (1.1) & 21.5\phn\ $\pm$ 0.9\phn\ (0.9)\phn\ & 0.27\phn\ $\pm$ 0.06\phn\ \\   
2008-Apr-01 & \Ks\    & 1.55 &  87 $\pm$ 4  & 0.095 $\pm$ 0.018        & 102.5 $\pm$ 0.7 (0.7) & 21.1\phn\ $\pm$ 0.7\phn\ (0.7)\phn\ & 0.25\phn\ $\pm$ 0.04\phn\ \\   
            & $CH_4s$ & 1.58 &  78 $\pm$ 7  & 0.048 $\pm$ 0.018        & 102.0 $\pm$ 0.4 (0.4) & 20.4\phn\ $\pm$ 1.5\phn\ (1.5)\phn\ & 0.21\phn\ $\pm$ 0.04\phn\ \\

\enddata

\tablenotetext{a}{All photometry on the MKO system.}
\tablenotetext{b}{The tabulated errors are computed by appropriately
  combining in quadrature: (1) the instrumental measurements from
  fitting the images of the binary and (2) the overall uncertainties
  in the NIRC2 pixel scale and orientation.  The errors in parentheses
  represent the instrumental errors alone.  See \S~2 for details.}

\end{deluxetable}


\begin{deluxetable}{lccccccc}
\tablecaption{\HST\ Observations \label{table:hst}}
\tablewidth{0pt}
\tablehead{
  \colhead{Date} &
  \colhead{Instrument} &
  \colhead{Filter} &
  \colhead{Separation\tablenotemark{a}} &
  \colhead{Position angle\tablenotemark{a}} &
  \colhead{$\Delta$mag} \\
  \colhead{(UT)} &
  \colhead{} &
  \colhead{} &
  \colhead{(mas)} &
  \colhead{(mas)} &
  \colhead{} 
}
\startdata
2000-Aug-18  & WFPC2  & $F814W$  & \phn62.8 $\pm$ 1.2  &  357.1 $\pm$ 0.8    &  0.30 $\pm$ 0.05  \\
2006-Jan-19  & ACS    & $F814W$  & 199.0 $\pm$ 1.1     & \phn14.5 $\pm$ 0.6  &  0.28 $\pm$ 0.06  \\
2006-Apr-11  & ACS    & $F814W$  & 191.2 $\pm$ 1.1     & \phn15.5 $\pm$ 0.4  &  0.30 $\pm$ 0.04  \\
\enddata

\tablenotetext{a}{The tabulated errors are dominated by the
  uncertainties in fitting the binary images, which are much larger
  than the errors in the overall astrometric calibration of WFPC2 and
  ACS.  See \S~2 for details.}

\end{deluxetable}

\clearpage
\begin{deluxetable}{lcc}
\tablecaption{Resolved Properties of \twomassbin\tablenotemark{a} \label{table:resolved}}
\tablewidth{0pt}
\tablehead{
  \colhead{Property} &
  \colhead{2MASS~J1534$-$2952A} &
  \colhead{2MASS~J1534$-$2952B}
}

\startdata
$F814W-J$ (mags) & \phn\phs4.95 $\pm$ 0.04 & \phn\phs 5.10$\pm$ 0.04  \\
$J-H$ (mags)     & \phn$-$0.08 $\pm$ 0.04  & \phn$-$0.21 $\pm$ 0.04  \\
$CH_4s-H$ (mags)\tablenotemark{c} & \phn$-$0.30 $\pm$ 0.05  & \phn$-$0.37 $\pm$ 0.05 \\
$H-K$ (mags)     & \phn$-$0.17 $\pm$ 0.04  & \phn$-$0.17 $\pm$ 0.04  \\
$J-K$ (mags)     & \phn$-$0.25 $\pm$ 0.04  & \phn$-$0.38 $\pm$ 0.04  \\
$M_{F814W}$ (mags)&  \phs19.57 $\pm$ 0.04  &  \phs19.87 $\pm$ 0.05  \\
$M_J$ (mags)     &  \phs14.61 $\pm$ 0.05  &  \phs14.77 $\pm$ 0.05  \\
$M_H$ (mags)     &  \phs14.69 $\pm$ 0.05  &  \phs14.98 $\pm$ 0.05  \\
$M_K$ (mags)     &  \phs14.86 $\pm$ 0.05  &  \phs15.15 $\pm$ 0.05  \\
Estimated spectral type\tablenotemark{b} & {\phn}T5.0 $\pm$ 0.5  &  {\phn}T5.5 $\pm$ 0.5 \\
$\log(\Lbol/\Lsun)$\tablenotemark{c}  & \phn$-5.015 \pm 0.019$  &  \phn$-5.093 \pm 0.019$  \\
\enddata

\tablenotetext{a}{All infrared photometry on the MKO photometric
  system.}  

\tablenotetext{b}{Based on the \citet{2005astro.ph.10090B} near-IR
  classification scheme.}

\tablenotetext{c}{The difference in the $(CH_4s-H)$ color of the two
  components is $0.07 \pm 0.02$~mags, \ie, better constrained than the
  quadrature sum of the measurement errors tabulated here.  See
  \S~\ref{sec:phot}.  Similarly, the difference in $\log(\Lbol/\Lsun)$
  is $0.078\pm 0.016$~dex, since this quantity is independent of the
  distance uncertainty.}

\end{deluxetable}

\clearpage
\begin{deluxetable}{lcccc}
\tablecaption{Derived Orbital Parameters for \twomassbin\ \label{table:orbits}}
\tabletypesize{\small}
\tablewidth{0pt}
\tablehead{
  \colhead{}                 &
  \multicolumn{3}{c}{MCMC}   &
  \colhead{ORBIT}            \\
  \colhead{}             &
  \colhead{Median}       &
  \colhead{68.3\% c.l.}  &
  \colhead{95.5\% c.l.}  &
  \colhead{} 
}

\startdata

Time of periastron $T_0-2400000.5$ (MJD)  &  55960\tablenotemark{a} &  $-$240, 210   &  $-$740, 450   &  56024 $\pm$ 347   \\   
Orbital period $P$ (yr)                   &   15.1      &  $-$1.6, 2.3        &  $-$3.1, 5.1          &   15.2 $\pm$ 2.6   \\
Semi-major axis $a$ (mas)                 &   171       &  $-$13, 19          &  $-$27, 41            &   172  $\pm$ 22    \\
Semi-major axis $a$ (AU)\tablenotemark{b} &   2.3       &  $-$0.2, 0.3        &  $-$0.4, 0.6          &   2.3  $\pm$ 0.3   \\
Inclination $i$ (\degs)                   &   84.3      &  $-$0.6, 0.6        &  $-$1.7, 1.0          &   84.3 $\pm$ 0.8   \\
Eccentricity $e$                          &   0.25      &  $-$0.13, 0.11      &  $-$0.20, 0.25        &  0.24 $\pm$ 0.16   \\
PA of the ascending node $\Omega$ (\degs) &   13.0      &  $-$0.3, 0.3        &  $-$0.9, 0.5          &  13.0 $\pm$ 0.4    \\
Argument of periastron $\omega$ (\degs)   &   179       &  $-$14, 6           &  $-$83, 11            &   178 $\pm$ 10     \\
Total mass (\Msun): fitted                &   0.0556    &  $-$0.0017, 0.0019  &  $-$0.004, 0.004      & 0.056 $\pm$ 0.004 \\
Total mass (\Msun): final                 &   0.056     &  $-$0.003, 0.003    &  $-$0.006, 0.007      & 0.056 $\pm$ 0.005 \\

\enddata

\tablenotetext{a}{February 3, 2012 UT.}

\tablenotetext{b}{Includes the uncertainty in the parallax and pixel scale.}

\tablecomments{Median values and confidence limits for orbital
  parameters derived from our default MCMC fitting, which uses a prior
  distribution flat in $\log(P)$ and $\log(a)$.  The ``fitted total
  mass'' represents the direct MCMC results from fitting the observed
  orbital motion of the two components.  The ``final total mass''
  includes the additional 4.9\% error from the uncertainties in the
  parallax and the Keck/NIRC2 pixel scale; the former is
  $\approx$15$\times$ larger than the latter.  The final mass
  distribution is essentially gaussian.  The rightmost column gives
  the results from the ORBIT routine by
  \citet{1999A&A...351..619F}. (See \S~\ref{sec:fitting}.)}

\end{deluxetable}

\clearpage

\begin{deluxetable}{l ll | ll | ll}
\tablecaption{Evolutionary Model-Derived Properties of \twomassbin\ \label{table:evolmodels}}
\tabletypesize{\footnotesize}
\rotate
\tablewidth{0pt}
\tablehead{
  \colhead{Property} &
  \multicolumn{2}{c}{Tucson models}  &
  \multicolumn{2}{c}{Lyon models}  &
  \multicolumn{2}{c}{``Average''}  \\
  \colhead{} &
  \colhead{Component A} &
  \colhead{Component B} & 
  \colhead{Component A} &
  \colhead{Component B} &
  \colhead{Component A} &
  \colhead{Component B}
}

\startdata
log(age)       & \multicolumn{2}{c|}{$\phn8.86\ _{-0.04(0.09)}^{+0.04(0.09)}$}                               & 
                 \multicolumn{2}{c|}{$\phn8.92\ _{-0.04(0.09)}^{+0.04(0.09)}$}                               &
                 \multicolumn{2}{c}{$\phn8.89\pm0.05(0.10)$}                                             \\[1.5ex]

Radius (\Rsun) & $\phn0.0997\ _{-0.0012(0.0024)}^{+0.0012(0.0025)}$  & $\phn0.1004\ _{-0.0012(0.0024)}^{+0.0012(0.0024)}$  & 
                 $\phn0.0978\ _{-0.0010(0.0022)}^{+0.0010(0.0021)}$  & $\phn0.0984\ _{-0.0012(0.0025)}^{+0.0011(0.0022)}$  &
                 $\phn0.0986\pm0.0015(0.0028)$                  & $\phn0.0993\pm0.0017(0.0031)$                  \\[1.5ex]

\Teff\ (K)     & $\phn\phn\phd1019\ _{-15(31)}^{+16(32)}$         & $\phn\phn\phn\phd970\ _{-15(29)}^{+15(31)}$           & 
                 $\phn\phn\phd1034\ _{-15(30)}^{+16(32)}$         & $\phn\phn\phd985\ _{-15(29)}^{+15(32)}$               &
                 $\phn\phd1028\pm17(35)$                       & $\phn\phn\phd978\pm17(34)$                          \\[1.5ex]

\logg\ (cgs)   & $\phn\phn\phn4.90\ _{-0.04(0.07)}^{+0.04(0.07)}$   & $\phn\phn\phn4.86\ _{-0.04(0.07)}^{+0.04(0.07)}$       & 
                 $\phn\phn\phn4.92\ _{-0.04(0.07)}^{+0.04(0.07)}$   & $\phn\phn\phn4.88\ _{-0.04(0.07)}^{+0.04(0.08)}$       &
                 $\phn\phn4.91\pm0.04(0.07)$                    & $\phn\phn4.87\pm0.04(0.07)$                        \\[1.5ex]

$\Delta$\Teff\ (K)  & \multicolumn{2}{c|}{$\phn\phn50\ _{-10(19)}^{+6(10)}$}                                  & 
                      \multicolumn{2}{c|}{$\phn\phn50\ _{-10(19)}^{+6(10)}$}                                  &
                      \multicolumn{2}{c}{$\phn 50\ _{-10(19)}^{+6(10)}$}                                                     \\[1.5ex]

Mass ratio $M_B/M_A$    & \multicolumn{2}{c|}{$\phn\phn0.937\ _{-0.007(0.012)}^{+0.012(0.024)}$}                                &
                          \multicolumn{2}{c|}{$\phn\phn0.934\ _{-0.008(0.013)}^{+0.012(0.025)}$}                                &
                          \multicolumn{2}{c}{$\phn\phn0.936\ _{-0.008(0.013)}^{+0.012(0.024)}$}                                       \\[1.5ex]

Mass (\Msun)   & $0.0287\ _{-0.0016(0.0032)}^{+0.0017(0.0036)}$     & $0.0270\ _{-0.0015(0.0030)}^{+0.0016(0.0033)}$ & 
                 $0.0288\ _{-0.0016(0.0031)}^{+0.0017(0.0035)}$     & $0.0269\ _{-0.0016(0.0030)}^{+0.0017(0.0034)}$ &
                 $0.0287\pm0.0016(0.0033)$                     & $0.0269\pm0.0016(0.0032)$  \\

\enddata

\tablecomments{Median values of physical parameters derived for the
  two components of \twomassbin\ from the evolutionary models along
  with the 68(95\%) confidence limits, as described in
  \S~\ref{sec:ages}.  By construction, the ages of the two components
  are identical for a given set of models, since the analysis assumes
  the system is coeval, and thus a single model-derived age is listed
  for both components.
  Note that the two sets of models predict nearly identical mass
  ratios so the two sets of individual masses are the same.}

\end{deluxetable}

\end{document}